\documentclass[twocolumn]{aastex7}

\usepackage{threeparttable}
\usepackage{rotating}

\begin{document}

\title{Disentangling Disk and Broad-Line Region Continuum Emission in a Sample of Active Galactic Nuclei}

\author[orcid=0009-0008-7583-5658]{Manqi Fu}
\affiliation{Department of Astronomy, Xiamen University, Xiamen, Fujian 361005, People’s Republic of China}
\email{fumanqi@stu.xmu.edu.cn}

\author[orcid=0009-0005-2801-6594]{Shuying Zhou}
\affiliation{Department of Astronomy, Xiamen University, Xiamen, Fujian 361005, People’s Republic of China}
\email{zhoushuying@stu.xmu.edu.cn}

\author[0000-0001-8416-7059]{Hengxiao Guo}
\affiliation{Shanghai Astronomical Observatory, Chinese Academy of Sciences, 80 Nandan Road, Shanghai 200030, People’s Republic of China}
\email[show]{hengxiaoguo@gmail.com}

\author[0000-0002-2419-6875]{Zhi-Xiang Zhang}
\affiliation{College of Physics and Information Engineering, Quanzhou Normal University, Quanzhou, Fujian 362000, People’s Republic of China}
\email{zzx@qztc.edu.cn}

\author[0000-0002-1380-1785]{Marcin Marculewicz}
\affiliation{Department of Physics and Astronomy, Wayne State University, 666 W. Hancock St, Detroit, MI 48201, USA}
\email{marcin.marculewicz@gmail.com}

\author[orcid=0000-0002-0771-2153]{Mouyuan Sun} 
\affiliation{Department of Astronomy, Xiamen University, Xiamen, Fujian 361005, People’s Republic of China}
\email[show]{msun88@xmu.edu.cn}

\begin{abstract}
Recent continuum reverberation-mapping measurements of active galactic nuclei (AGNs) indicate that the inter-band time lags are significantly larger than those predicted by the light travel times across the standard thin model. In addition, the observed $u$-band excess suggests the presence of diffuse continuum (DC) radiation originating in the broad-line region (BLR). Spectroscopically, the Balmer and Paschen jumps in DC emission are unique indicators to quantify the contributions of the BLR to the observed continuum. Here, we study the DC contributions of 31 AGNs, whose spectra are obtained with the Space Telescope Imaging Spectrograph (STIS) on the \textit{Hubble Space Telescope} (\textit{HST}) and with X-shooter on the Very Large Telescope (VLT). In each spectrum, at least one of the Balmer and Paschen jumps is covered. For each object, we performed radiative transfer calculations using \texttt{Cloudy} within the Locally Optimally-emitting Clouds (LOC) model to construct source-specific DC spectral templates. We quantified the DC contribution by incorporating these templates as continuum components in spectral decomposition. The average ratio of the DC to total (i.e., disk+DC) continua is $\simeq 33\%$ just below the Balmer jump ($\sim 3625$\,\AA). The Balmer jump provides a stronger constraint on the DC component than the Paschen jump. Simulations based on mock spectra further demonstrate that an unsuitable template can introduce systematic, wavelength-dependent biases to the inferred DC fraction. Our results underscore the importance of employing appropriate DC templates in the spectral decomposition to infer the substantial DC contribution. Upcoming time-resolved spectra will enable more detailed modelling and tighter constraints on the disk and BLR.

\end{abstract}


\section{Introduction}\label{sec:intro}
It is generally accepted that active galactic nuclei (AGNs) are powered by accreting supermassive black holes (SMBHs); other important components are broad-line regions (BLRs), dusty tori, and narrow-line regions \citep[for a recent review, see, e.g.,][]{AGN_Unified}. The innermost regions of the accretion disk emit soft X-ray and extreme ultraviolet radiation, which photoionize the high-velocity BLR clouds in the vicinity of the SMBH, producing broad emission lines. Due to the extremely small angular scales of SMBH accretion disks and their BLR clouds, it is challenging to resolve their spatial structures directly with current observational facilities. Indeed, accretion disks have only been resolved in Sgr~A$^*$ \citep{Akiyama_2023} and M~87$^*$ \citep{Akiyama_2019}, while BLRs have been resolved in several AGNs
(e.g., \citealp{Gravity_2018} for 3C~273; \citealp{Gravity_2026} for SMSS~J052915.80-435152.0).

Currently, two primary indirect methods based on temporal information are used to probe the scale of accretion disks: the microlensing effect of gravitationally lensed quasars \citep{Chang_1979,Kochanek_2004} and continuum reverberation mapping \citep{Cackett_2007}. However, the sizes of accretion disks obtained from both methods are significantly larger than the theoretical predictions of the standard disk model \citep[e.g.,][]{Morgan_2010,McHardy2014,Fausnaugh_2016,Fian_2021,Guo_2022rm,Cackett_2023,Fores_2024,zhou_2025,Feng_2026,Miller_2026,Marculewicz_2026}, suggesting the presence of physical mechanisms not yet fully understood. To address the larger-than-expected accretion disk problem, a series of theoretical modifications to the standard disk have been proposed, such as non-blackbody emission from scattering in a low-density atmosphere \citep{Hall_2018}, disk winds \citep[e.g.,][]{Sun_2018, Li_2019}, inhomogeneous accretion disks \citep[e.g.,][]{Jason_2011}, and intrinsic radially coherent disk variations \citep[e.g.,][]{Cai_2018} possibly due to magnetic coupling with the standard thin disk and the corona \citep{Sun_2020}. 

In continuum reverberation mapping, the time delays observed in the light curves of different bands are often interpreted as the travel time for light from the X-ray corona to propagate to different radiation regions \citep[e.g.,][]{Cackett_2007}. Consequently, the time delays effectively reflect the distance between the radiation region and the X-ray corona. It is proposed that, in addition to the continuum emission from the standard disk, a diffuse continuum (DC) component from the BLR, located farther from the central black hole, can also significantly enlarge the continuum time-lag measurements \citep[e.g.,][]{Cackett_2018, Korista2019}. The recent studies by \citet{lawther2018} and \citet{Korista2019} assessed the flux and lag spectra of DC emission and found that DC emission in NGC 5548 can contribute $\sim$40\% to the total continuum emission at wavelengths around the Balmer jump. Significant DC contributions have also been quantified in several AGNs \citep[e.g.,][for Mrk 817]{Netzer_2024}. There remains uncertainty regarding the DC contribution in most AGNs, even whether it universally exists. It is also important to estimate the DC contributions in a large AGN sample. 

DC radiation exhibits the distinctive features of the Balmer jump (3646~\AA) and the Paschen jump (8206~\AA). Continuum reverberation mapping campaigns have identified a distinct excess in lag within the $u$-band spectral region in several objects, which has been attributed to this DC emission \citep[e.g.,][]{Edelson_2015, Fausnaugh_2016, Cackett2018, Cackett2020, Hernandez2020, Feng_2026}. Additionally, the discrepancy between theoretical predictions of AGN accretion disk sizes and gravitational microlensing observations can be partially resolved by accounting for DC radiation \citep[e.g.,][]{Fian_2023}. 

The distinctive spectroscopic Balmer and Paschen jumps provide key indicators for identifying the DC contribution to the observed AGN continua. \citet{Guo_2022} performed multi-component decomposition on near-infrared spectra of five nearby AGNs utilizing the Paschen jump, suggesting that the DC contributes approximately 10\%--50\%. This result provides new evidence for the DC contribution in the infrared band and demonstrates the potential feasibility of spectral decomposition methods for quantitatively assessing DC radiation. Quantitatively assessing the contribution of DC to the AGN continuum via the spectral decomposition technique not only helps interpret time-domain accretion disk observations correctly and constrain their actual structure, but also provides crucial evidence for refining accretion disk reprocessing models and deepens our understanding of the physical properties of the BLR.

Here, we extend the work of \citet{Guo_2022} by systematically analyzing the UV-to-NIR spectra of 31 AGNs obtained with the Space Telescope Imaging Spectrograph \cite[STIS;][]{Kimble_1998} on the \textit{Hubble Space Telescope} (\textit{HST}) and X-shooter \citep{Vernet_2011} on the Very Large Telescope (VLT). These data were used to construct dedicated DC templates and to perform multi-component spectral fitting, thereby constraining the DC contribution. The structure of the manuscript is as follows. In Section~\ref{sec:data}, we describe the sample selection and spectral decomposition procedure. In Section~\ref{sec:DC template}, we present the construction of the DC templates based on the Locally Optimally-emitting Clouds (LOC) model \citep{Baldwin_1995}. In Section~\ref{sec:res}, we report the resulting DC templates and spectral decomposition results, along with a discussion of their implications. Finally, Section~\ref{sec:conclusion} summarizes our main conclusions. We adopt the flat $\Lambda$CDM cosmology with $H_0 = 70~\mathrm{km\,s^{-1}\,Mpc^{-1}}$, $\Omega_{\mathrm{M}} = 0.3$, 
and $\Omega_{\Lambda} = 0.7$, unless otherwise specified. 

\section{Observation data and decomposition method} \label{sec:data}
\subsection{Sample Selection}

Our sample was drawn from the updated catalog of Broad-Line Reverberation-Mapped (BLRM) AGNs \citep{Wang_2024} and from the ongoing Intensive Broadband Reverberation Mapping AGN sample in the AGN Variability Archive (AVA) of the LCO Key Project.\footnote{\url{https://alymantara.com/ava/index.php}} There are 78 sources in AVA and 212 AGNs in the BLRM AGN catalog. For each source, we searched the \textit{HST} archive\footnote{\url{https://mast.stsci.edu/search/ui/\#/hst}} for STIS observations with the G140L (1150--1730~\AA), G230L (1570--3180~\AA), G430L (2900--5700~\AA), and G750L (5240--10270~\AA) gratings, and the VLT archive\footnote{\url{https://archive.eso.org/scienceportal/home}} for X-shooter spectra spanning 3100--5500~\AA\ (UVB arm) and 5500--10200~\AA\ (VIS arm). These two instruments have moderate to high spectral resolutions ($R\sim3000-18000$ for X-shooter and $R\sim1000-3000$ for the STIS low-resolution gratings), and their wide wavelength ranges cover both the Balmer and Paschen jumps for low-redshift objects. The maximum search radius was 5 arcsec. In this process, we yielded 39 Seyfert 1 galaxies (21 from STIS and 18 from X-shooter) after removing NGC 3227 and IC 4329A, whose spectra are severely affected by host-galaxy extinction. We also rejected NGC 7469, as this target will be explored in detail by Marculewicz et al. in preparation. Hence, the sample contains 38 sources.

\subsection{Data reduction}\label{sec:reduction}
We downloaded the raw STIS data and calibration files and processed them for cosmic-ray removal and charge-transfer-inefficiency (CTI) corrections using the \texttt{stistools.calstis} package from \texttt{stistools}.\footnote{\url{https://stistools.readthedocs.io/en/latest/index.html}} The G750L spectra at wavelengths longer than 7000 \AA\ are affected by ``fringing'' caused by interference of multiple reflections between the two surfaces of the CCD \citep{Goudfrooij_1999}. Following the standard procedures, we obtained the appropriate flat-field correction files from \textit{HST} archive and applied them to remove fringing using the \texttt{stistools.defringe} package. To further improve cosmic-ray removal, we applied a consistent continuum window (20 pixels) and excluded data points whose deviations from the mean exceeded $3\sigma$ after masking out emission lines. For each source with multi-epoch STIS spectra, we compared spectra across epochs and excluded data points at wavelengths where the difference between the maximum and minimum fluxes exceeded five times the mean flux. After this cleaning procedure, we averaged the spectra from different epochs to obtain the final STIS spectrum for each source. An exception is Mrk 817, for which we selected the epoch with the highest data quality (the highest average data-to-error ratio), as most of its spectra are significantly affected by fringing that cannot be effectively removed. Obvious UV absorption lines were additionally masked in the final spectra prior to fitting. 

We also retrieved the X-shooter spectra. For each target, we selected the spectrum with the highest signal-to-noise ratio (SNR) among the observations. The selected spectra were subsequently corrected for telluric absorption by modeling atmospheric absorption spectra. To determine the optimal telluric model, we generated high-resolution telluric transmission spectra using \texttt{Telfit}\footnote{\url{https://telfit.readthedocs.io/en/latest/}}\citep{Gullikson_2014} and fitted the observed spectra by minimizing the residuals between the data and the model using \texttt{spectool}.\footnote{\url{https://github.com/zhang-zhixiang/spectool}} The model was defined as the product of the telluric transmission and a continuum component, represented by a low-order polynomial function. The best-fit transmission model was then applied to correct the observed spectra, yielding the intrinsic source flux. Prior to the subsequent spectral analysis, we masked the observed-frame 7580--7680~\AA\ region. This range is heavily contaminated by the prominent telluric $\mathrm{O_2}$ A-band \citep{Smette_2015}, where the deep absorption features often leave significant residuals after telluric correction.

\subsection{Object properties}\label{sec:information}
The black hole masses and Eddington ratios adopted in our analysis are primarily taken directly from the Swift BAT AGN Spectroscopic Survey \citep[BASS DR2;][]{Koss2022}. For sources not included in BASS DR2, we obtained their black hole masses ($M_{\mathrm{BH}}$) and bolometric luminosities ($L_{\mathrm{bol}}$) from the AGN Black Hole Mass Database \citep{Bentz_2015} or other literature listed in Table~\ref{tab:obs_t}. The Eddington ratios are calculated as $\dot{m} = L_{\mathrm{bol}}/L_{\mathrm{Edd}}$, where $L_{\mathrm{Edd}} = 1.26\times10^{38}\,(M_{\mathrm{BH}}/M_\odot)\,\mathrm{erg\,s^{-1}}$. The observational and source information is also shown in Table~\ref{tab:obs_t}. In Section~\ref{subsec:Parameter determination}, we removed seven sources that did not meet the corresponding line-ratio criteria; hence, this table includes only the 31 sources that were ultimately subjected to spectral decomposition.

\begin{deluxetable*}{lccllccccl}
\tablenum{1}
\tablecaption{Sample information and observation information \label{tab:obs_t}}
\tablehead{
\colhead{Object}&\colhead{Instrument} &\colhead{Obs. date} &\colhead{PID} &\colhead{PI last name}& \colhead{$z$} & \colhead{$\log_{10}(\mathit{M_\mathrm{BH}/\mathit{M}_\odot})$} & \colhead{$\log_{10}\mathit{\dot{m}}$} & \colhead{Ref.} & \colhead{$\lambda_{\mathrm{rest}}(\mathrm{\AA})$}}
\colnumbers{}
\startdata
ESO 362-G18&VLT/X-shooter & 2022~Nov~01  &110.241X& Trakhtenbrot &0.012 & 7.11 &	-1.22& (a) &[2953, 10074]	\\
ESO 511-30&VLT/X-shooter   & 2022~Jan~17&108.229H&Trakhtenbrot&0.022 &	7.23	& -0.88	& (a)&[2923, 9971]\\
Fairall 9 &VLT/X-shooter &  2013~Sep~11 &091.B-0900&Vestergaard &0.046 &	8.30&	-1.23& (a) &[2863, 9753]\\
H 0507+164&VLT/X-shooter & 2021~Dec~14  &108.229H&	Trakhtenbrot&0.017	&6.88&	-0.41	& (a)&[2938, 10026]\\
Mrk 590	&VLT/X-shooter  & 2017~Dec~24 &2100.B-5018&Vestergaard&0.026&7.57&	-1.51	& (a)&[2912, 9937]\\
Mrk 1044	&VLT/X-shooter  &2019~Oct~30  &0104.B-0959&Urrutia&0.016 &6.45	&-0.90	& (a)&[2938, 10027]\\
Mrk 1048	&VLT/X-shooter & 2019~Oct~31  &0104.B-0959&Urrutia&0.043&8.11	&-1.32	& (a)&[2866, 9779]\\
Mrk 1239  &VLT/X-shooter   & 2021~Dec~01&108.23LY&Shangguan&0.019	&6.81	&-0.36	& (b) &[2933, 10006]\\
Mrk 1298	&VLT/X-shooter  & 2020~Feb~16 &0104.B-0959&Urrutia&0.059	&7.75	&0.42	&(c) &[2821, 9624]\\		
NGC 2617	&VLT/X-shooter  & 2021~Oct~27 &108.229H&Trakhtenbrot&0.014	&6.93&	-1.96& (a)&[2948, 10057]\\
NGC 3783&VLT/X-shooter  & 2013~Feb~11 &090.A-0830&Brusa	&0.009	&7.37&	-1.20	& (a)&[2963, 10109]\\
NGC 4748	&VLT/X-shooter & 2023~Mar~04  & 110.241X&Trakhtenbrot&0.014	&6.41	&-0.96	&(a)&[2948, 10058]\\
PG 2214+139	&VLT/X-shooter  & 2019~Aug~01 &0103.A-0521&Trakhtenbrot&0.065	&8.17	&-1.49& (a)&[2806, 9573]\\
PG 0934+013 &VLT/X-shooter  &2020~Feb~16   &0104.B-0959	&Urrutia&0.050	& 6.5	&-0.44	& (d)&[2848,9712]	\\
PKS 2004-447&VLT/X-shooter  &2019~Nov~25  &104.20UC	&Berton&0.240	&7.60	&-0.25	& (e) &[2411, 8226]	\\
3C 382&\textit{HST}/STIS  & 2003~Jul~23  & 9495 &	Martel&0.055&	8.01&	-0.56& (a)&[2740, 9687]\\
Arp 151 &\textit{HST}/STIS&2013~Apr~29&12922&Woo	&0.020	&6.67	&-0.73	& (a)&[1115, 5581]\\
Mrk 50&	 \textit{HST}/STIS  & 2012~Dec~12 &12922&Woo&0.023&7.42&	-1.29& (a)&[1112, 5566]\\
Mrk 279&\textit{HST}/STIS  & 2023~Jun~07-15 &16705&	Chelouche &0.030&	7.43	&-0.86	& (a)&[1105, 9932]\\
Mrk 817	 &\textit{HST}/STIS  &2022~Feb~24  & 16196&Kriss&0.031	&7.59&	-1.16&(a)&[1104, 9940]\\
Mrk 1310	&\textit{HST}/STIS & 2013~Jun~07   &12922&Woo&0.019&	6.21&	-0.57& (a)&[1116, 5588]\\
NGC 6814	&\textit{HST}/STIS & 2013~May~07  &12922&Woo&0.005&	7.04&	-1.70& (a)&[1132, 5665]\\
PG 1226+023&\textit{HST}/STIS&  1999~Jan~31  &7568	&Hutchings&0.158&	8.84&	0.30	& (a)&[2505, 8862]\\
PHL 1811&\textit{HST}/STIS &  2001~Dec~03 &9181	&Leighly&0.194	&8.26&	0.09& (f) &[953, 4776]\\
SBS 1116+583A	&\textit{HST}/STIS& 2013~Jul~12   &12922&Woo&0.027	&6.78	&-0.08& (g)&[1108, 5543] \\
ZW 229-015&\textit{HST}/STIS & 2013~Jul~23 &12922&Woo&	0.027	&6.91	&-0.99& (a)&[1107, 5543]\\
Mrk 110 &\textit{HST}/STIS   &2017~Dec~25-2018~Jan~10 &15413	&Cackett&0.035&	7.29	&-0.41& (a)&[1526, 9885]\\ 
Mrk 493	 &\textit{HST}/STIS  &2017~Aug~28-31  &14744&Park&0.031&	6.17	&0.16& (h) &[1104, 9925]\\		
Mrk 509	&\textit{HST}/STIS   &2017~Oct~22 &15124&Barth&0.034	&8.05&	-0.96	& (a)&[1100, 9887]\\			
NGC 4151 &\textit{HST}/STIS  &2000~May~24-28  &8136&Mundell&	0.003&	7.56&	-1.53& (a)&[1135, 10236]\\		
NGC 4593 &\textit{HST}/STIS  &2016~Jul~12-Aug~6  &14121 &Cackett&	0.008&	6.88&	-1.05& (a)&[2864, 10153]\\	
\enddata

\tablecomments{Column (1): Object name. Column (2): Spectrum source (telescope and instrument). Column (3): Observation date. Column (4): Proposal ID for HST observations and program ID for VLT observations. Column (5): Last name of the principal investigator (PI). Column (6): Redshift. Column (7): Black hole mass. Column (8): Eddington ratio. Column (9): References for black hole mass and bolometric luminosity: (a) BASS DR2; (b) \citet{Doi_2015}; (c) \citet{Vestergaard_2006}; (d) \citet{Park_2017}; (e) \citet{Berton2021}; (f) mass from \citet{Leighly_2007} and bolometric luminosity from \citet{Vestergaard_2006} and \citet{Netzer_2019}; (g) \citet{Liu_2021}; (h) \citet{Bentz_2015}. Column (10): Rest-frame wavelength coverage.}
\end{deluxetable*}

\subsection{Spectrum fitting method} \label{subsec:Spectrum fitting}

We performed spectral fitting using the code \texttt{PyQSOFit} \citep{pyqsofit_2018, Shen_2019}, incorporating modified template components to model the continuum emission. The continuum model includes contributions from the accretion disk, the DC component, Fe\,{\sc ii} emission, host-galaxy starlight, the high-order Balmer and Paschen line series, and dust emission:
\begin{equation}
f_{\mathrm{conti}} = f_{\mathrm{disk}} + f_{\mathrm{DC}} + f_{\mathrm{Fe}} + f_{\mathrm{host}} + f_{\mathrm{b/p}} + f_{\mathrm{dust}}.
\end{equation}
Our fitting covers the wavelength range from the near-ultraviolet to the near-infrared (for each source, the rest-frame wavelength coverage of the fitting is shown in Table~\ref{tab:obs_t}). The optimal parameters were determined by minimizing the $\chi^2$ after masking prominent emission lines. 

The DC template for each source was computed using \texttt{Cloudy} \citep{Ferland_1998,cloudy_2025} within the framework of the LOC model \citep{Baldwin_1995} for the BLR, as described in detail in Section~\ref{sec:DC template}. The free parameters of the DC component include velocity broadening, a normalization factor, and a wavelength offset, with the template allowed to vary by up to five pixels (about 15~\AA) to account for minor alignment uncertainties. The spectral velocity broadening was achieved by convolving the DC spectrum with a Gaussian kernel after interpolating it onto a uniformly spaced logarithmic wavelength grid. The broadened spectrum was then re-interpolated onto the wavelength grid of the observed data. In addition, we tied the DC velocity broadening to that of the Balmer series, assuming both originate from the same region, thereby constraining the velocity of the DC component.

The high-order Balmer and Paschen line series were generated using the Case B recombination emissivities of \citet{Hummer_1987}. The predicted line ratios up to $n=50$ were adopted directly and extrapolated to $n=164$ using a polynomial fit. The resulting line spectra were then convolved with the same velocity broadening applied to the DC. Further details of the construction of the Balmer and Paschen series were given in \citet{Guo_2022}. Each high-order line series was characterized by three free parameters: velocity broadening, normalization, and wavelength offset.

We modeled the accretion disk continuum as a broken power law, motivated by the prominent spectral break observed between 4000 and 6000~\AA\ in the SDSS composite quasar spectrum \citep{Berk_2001}. The short- and long-wavelength spectral indices were both constrained to be non-positive ($\alpha_{\rm short} \leq 0$ and $\alpha_{\rm long} \leq 0$), with the short-wavelength spectral index further constrained to the range $[-5, 0]$. The break wavelength ($\lambda_{\rm break}$) was treated as a free parameter varying between 3000 and 9000~\AA. To ensure a flatter continuum at longer wavelengths, we imposed the condition $|\alpha_{\rm short}| \ge |\alpha_{\rm long}|$ (i.e., the short-wavelength slope is steeper than or equal to the long-wavelength slope). The broken power-law model naturally reduces to a single power law when $\lambda_{\rm break}$ lies outside the fitted spectral range or when $\alpha_{\rm short} = \alpha_{\rm long}$. Modeling the disk as a broken power law is a highly simplified approach. Therefore, we further tested the potential impact of using different accretion disk models on the fit results for Mrk~1048 in Appendix~\ref{disktest}.

The template of Fe\,{\sc ii} emission complexes came from the built-in template from \texttt{PyQSOFit}, which contains the UV Fe\,{\sc ii} component from 1200 to 3500~\AA\ \citep{Vestergaard_2001}, and the optical Fe\,{\sc ii} component from 3686 to 7484~\AA\ \citep{Boroson_1992}. The free parameters of each Fe\,{\sc ii} component also include velocity broadening, a normalization factor, and a wavelength offset.

The hot dust emission was modeled using a single-temperature blackbody component, with the temperature allowed to vary from 1000 to 2000~K and the normalization treated as a free parameter.

For sources without a significant host-galaxy contribution, the continuum decomposition involves the 20 free parameters described above, whereas for sources with a non-negligible host-galaxy contribution, we additionally adopted extinction-modified templates based on the BC03 models \citep{BC03} to fit the host-galaxy stellar component. The BC03 models provide single stellar population (SSP) spectra spanning a broad range of ages ($10^5$~yr to 20~Gyr) and metallicities ($Z = 0.0001$--$0.05$), and predict high-resolution optical spectra for each combination of age, metallicity, and initial mass function (IMF). To account for internal dust attenuation, we applied the Calzetti extinction law \citep{Calzetti_2000} to the 39 BC03 templates using 9 color excess values $E(B-V) = 0.05,~0.1,~0.15,~0.2, ~0.25,~ 0.3,~ 0.35,~0.4 $ and $0.5$. Including the original unreddened templates, this process resulted in a final library of 390 galaxy spectra. Following the fitting procedure in \texttt{PyQSOFit}, we used these 390 galaxy spectra together with 10 AGN principal component analysis (PCA) templates \citep{Yip_2004} in the first stage of spectral fitting, allowing the normalizations of all 400 templates to vary freely. The best-fitting host-galaxy component was determined in this initial stage and then subtracted from the observed spectrum. The resulting host-subtracted spectrum was used for the subsequent continuum decomposition. For sources without a significant host-galaxy contribution, no host-galaxy subtraction was required.

\section{Diffuse Continuum spectral template construction} \label{sec:DC template}

\subsection{LOC model} \label{subsec:LOC model}

\begin{figure} 
\includegraphics[width=1. \linewidth]{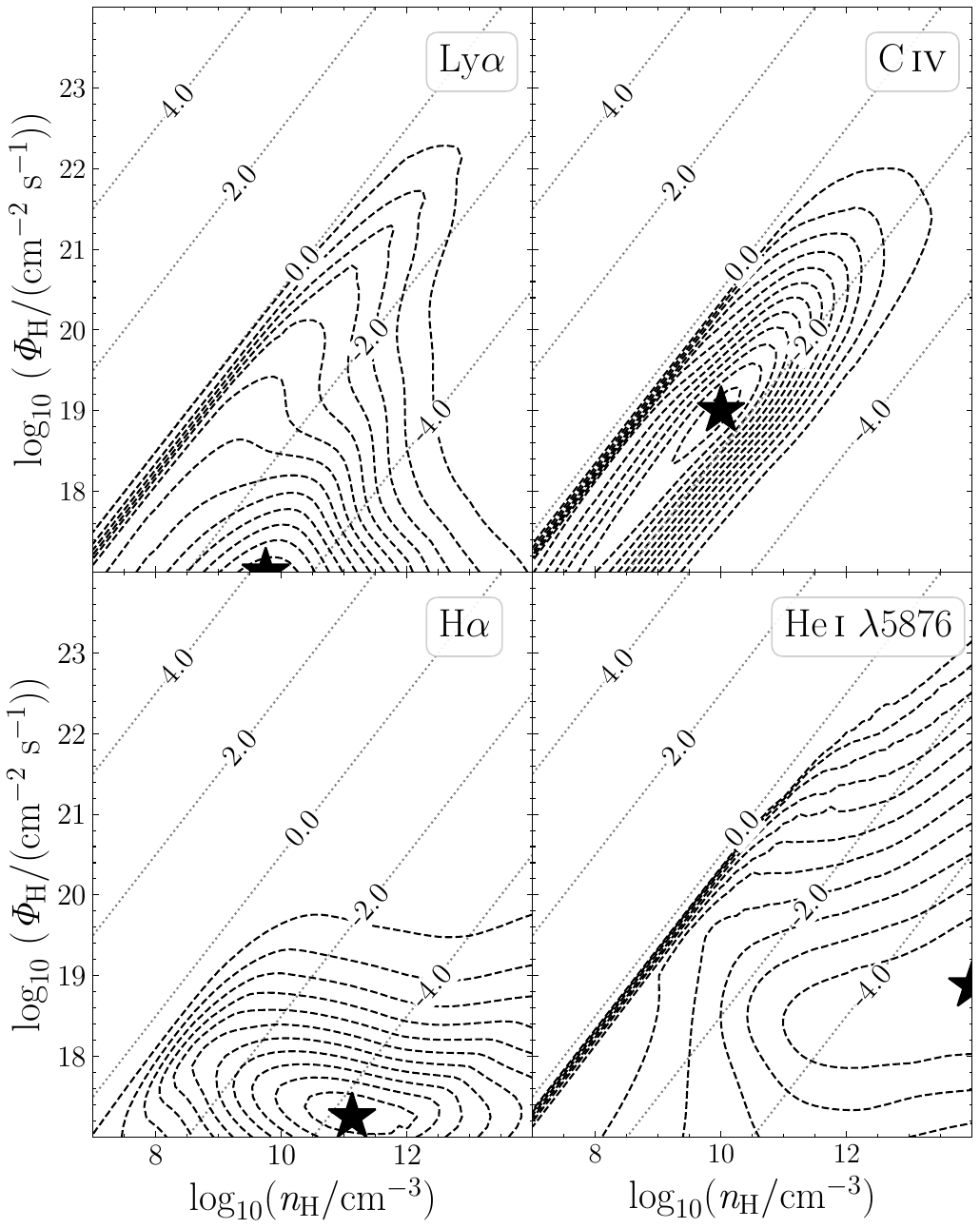}
\caption {Logarithmic contours of the EW for Ly$\alpha$, C\,\textsc{iv}, H$\alpha$, and He\,\textsc{i}~$\lambda5876$ in the gas hydrogen number density--incident ionizing photon flux plane for photoionized clouds with $N_{\rm H}=10^{23}\rm cm^{-2}$. The black star marks the location of the maximum EW, and the contours decrease in intervals of 0.2 dex (dot-dashed lines). The gray dashed lines indicate constant ionization parameters ($U_\mathrm{H}$). The peak EWs occur in distinct regions. 
Hence, line ratios can be used to constrain the spatial distribution of the BLR.}
\label{cloudy}
\end{figure}

We modeled the BLR photoionization grid using \texttt{Cloudy} v17.01 \citep{Ferland_1998,cloudy_2025} and generated DC spectra for each source within the LOC framework \citep{Baldwin_1995}. The LOC model provides a reasonable explanation for both the observed broad emission-line ratios and the different spatial distributions of high- and low-ionization emission lines.

We adopted the grid of photoionization computations presented in Section 2.1 of \citet{Korista_2000} to model the BLR. Our LOC grid assumed a fixed column density of
$N_{\mathrm{H}} = 10^{23}\,\mathrm{cm^{-2}}$, and spanned a constant hydrogen number-density range of $8 \le \log_{10}(n_{\mathrm{H}}/\mathrm{cm^{-3}}) \le 12$, together with an ionization parameter range of $6 \le \log_{10}(U_{\mathrm{H}}c/(\mathrm{cm\,s^{-1}})) \le 11.25\,$ (where $U_{\mathrm{H}}$ is the dimensionless hydrogen ionization parameter, $c$ is the speed of light, and $U_{\mathrm{H}}\mathit{c}= \mathit{\Phi}_\mathrm{H}/\mathit{n}_\mathrm{H}$, where $\mathit{\Phi}_\mathrm{H}$ is the incident ionizing photon flux), which were consistent with the setups in \citet{Korista2019}. These setups cover a reasonable and effective range of emission parameters for BLR clouds \citep{Korista_1997}.

We modeled the spectral energy distribution (SED) of NGC~5548 from \citet{NGC5548_SED} using \texttt{AGNvar}\footnote{\url{https://github.com/scotthgn/AGNvar}} \citep{AGNvar}, obtaining a representative set of accretion and corona parameters (e.g., photon index, size, and electron temperature in different regions). This fiducial model consists of three regions, with the absolute size scale set by the black hole mass via the Schwarzius radius $R_{\mathrm{g}}$. The innermost region ($R<20$~$R_{\mathrm{g}}$) contains a hot corona with an electron temperature of 50~keV, producing a hard Comptonized power-law continuum with photon index $\mathit{\Gamma}_\mathrm{h} = 1.7$. At the radii of 20--60~$R_{\mathrm{g}}$, a warm Comptonizing region with electron temperature of 0.23~keV, yields a photon index $\mathit{\Gamma}_\mathrm{w} = 2.1$. Beyond $60~R_{\mathrm{g}}$, there is a standard disk extending out to $10^5~R_{\mathrm{g}}$, emitting like a multi-temperature blackbody. These parameters were then adopted as the baseline configuration. For each source, we generated a specific SED using its black hole mass and Eddington ratio, and employed these SEDs as the incident ionizing continua in the subsequent \texttt{Cloudy} calculations.

Given the incident SED, the incident ionizing photon flux, $\mathit{\Phi}_\mathrm{{H}}$, depends on the radial distance of the BLR clouds, $R$, and is defined as: 
\begin{equation}\label{eq:phiout}
\mathit{\Phi}_\mathrm{{H}}~(\mathrm{cm^{-2}~s^{-1}}) = \frac{\mathit{Q}_\mathrm{H}}{4\pi \mathit{R}_{\mathrm{}}^2 } \\,
\end{equation}
in which $\mathit{Q}_\mathrm{H}$ is the total ionizing photon emission rate ($E_{\mathrm{ph}} > 13.6$~eV, in photons~s$^{-1}$) integrated from the SED.

Figure~\ref{cloudy} shows the equivalent width (EW), defined as $F_{\mathrm{line}} / F_{\lambda}^{\mathrm{inc}}$, where $F_{\mathrm{line}}$ is the emission-line flux and $F_{\lambda}^{\mathrm{inc}}$ is the incident continuum flux density at the line wavelength, for four representative broad emission lines (Ly$\alpha$, C\,\textsc{iv}, H$\alpha$, and He\,\textsc{i}~$\lambda5876$) in the gas hydrogen number density--incident ionizing photon flux plane for photoionized clouds with a fixed hydrogen column density of $N_{\mathrm{H}} = 10^{23}\,\mathrm{cm^{-2}}$. In each panel, the location of the maximum EW and the lines of constant ionization parameter ($\mathit{U}_\mathrm{H}$) are indicated. Different emission lines reach their maximum emissivities under different physical conditions, reflecting their distinct dependencies on gas density and incident ionizing photon flux. Consequently, observed emission-line ratios provide valuable constraints on the physical conditions and cloud distribution within the BLR. 

In the LOC model, the BLR consists of a large number of clouds with various densities and covering factors. The total continuum or line luminosity predicted by the LOC model can be computed by integrating over the distributions of cloud hydrogen number density and radial distance from the ionizing source:
\begin{equation}
L_{\mathrm{\lambda}} = 
A\int_{}
\int_{R_{\mathrm{in}}}^{R_{\mathrm{out}}}
4\pi r ^2F_{\mathrm{\lambda}}\!\left(n, r{\mathrm{}}\right)
\, g(n)C(r) \,
\, \mathrm{d}n_{\mathrm{}} \, \mathrm{d}r_{\mathrm{}} ,
\end{equation}
where $F_{\mathrm{\lambda}}(n,r)$ is the diffuse continuum or line emissivity from the photoionization calculations, and $g(n)$ and $ C(r)$ describe the cloud distributions in gas hydrogen number density and radius, respectively; $R_\mathrm{in}$ and $R_\mathrm{out}$ represent the inner and outer radii of the BLR clouds, and $A$ is the normalization constant ensuring that the total covering factor of all clouds completely obscures the ionizing source, calculated as:
\begin{equation}
\frac{1}{A}= 
\int_{}
\int_{R_{\mathrm{in}}}^{R_{\mathrm{out}}}
\, g(n)C(r) \,
\, \mathrm{d}n_{\mathrm{}} \, \mathrm{d}r_{\mathrm{}}.
\end{equation}

We used one-tenth of the BLR distance measured from the $\mathrm{H}\beta$ time delay as the inner radius (i.e., $R_{\mathrm{in}}$) of the BLR. The $\mathrm{H}\beta$ time delays were from the observational measurements of \cite{Wang_2024}; for sources with multiple measurements, we chose the one with the closest observational epoch to our spectral observation epoch. For sources that currently lack broad line reverberation mapping measurements, we used the 5100~\AA\ luminosity to estimate the expected $\mathrm{H}\beta$ radius from the $R$--$L$ relation \citep{Bentz_2013}. The 5100~\AA\ luminosities were either adopted from BASS DR2 or derived from the bolometric luminosities provided by AVA, assuming $L_{5100\,\text{\AA}} = 0.1\,L_{\mathrm{bol}}$ \citep{Richards_2006}.

We used the empirical density distribution $g(n) \propto n^{-1}$ \citep{Baldwin_1997}. Hence, we set the covering factor power-law index $s$, which describes the covering factor as a function of radius, $C(r) \propto r^{s},$ and the minimum incident ionizing photon flux\footnote{Note that $\mathit{\Phi}_{\mathrm{out}}$ and $R_{\mathrm{out}}$ are related via Eq.~\ref{eq:phiout}} $\mathit{\Phi}_{\mathrm{out}}$ of the cloud as free parameters. 

For gas metallicity, we adopted the metallicity prescription established for NGC~5548 from \cite{Korista2019}: the carbon and nitrogen abundances are set to solar values, while all other elements are fixed at 0.5 solar metallicity. To improve the accuracy of the recombination spectrum, we also adopted a larger number of resolved and collapsed levels for hydrogen and helium than the default settings in \texttt{Cloudy}. For hydrogen, all levels up to a principal quantum number of $n = 18$ are treated with fully resolved angular momentum states, while levels from $n = 19$ to $48$ are collapsed the angular momentum states into single effective levels. For neutral and singly ionized helium, the corresponding limits for resolved and collapsed levels are $n = 15$ and $55$, and $n = 15$ and $45$, respectively. And we adopted a relative optical-depth convergence criterion of 0.01 in Cloudy, requiring relative optical depths to change by less than 0.01 between the last two iterations.

\subsection{Parameter determination and Template Generation} \label{subsec:Parameter determination}

The free model parameters (i.e., $\mathit{\Phi}_{\mathrm{out}}$ and $s$) were constrained by modeling the observed emission-line intensity ratios. Selected emission-line ratios were measured by using \texttt{PyQSOFit} to fit the \textit{HST} or VLT spectra in the in the $1000$--$3000$~\AA\ and $4000$--$8000$~\AA\ ranges, respectively. In this step, we did not use the DC component, as we focused on emission line ratios. 

To minimize the impact of dust extinction on these ratios, we divided the lines into ultraviolet and optical subsets, separated at the Balmer jump. We used the built-in AGN table SED in \texttt{Cloudy} to explore the metallicity dependency. We computed ultraviolet and optical emission-line ratios from LOC models spanning metallicities from $0.5Z_\odot$ to $5Z_\odot$. From the \texttt{Cloudy} calculation, we identified emission lines whose mean variations remained within $\pm~15\% $ across the explored metallicity range and therefore exhibited only weak sensitivity to metallicity. Restricting our analysis to these lines helps mitigate potential systematic uncertainties arising from metallicity variations in the subsequent fitting procedure. Nine emission-line ratios were retained, including 
H$\beta$/H$\alpha$, He\,\textsc{ii}$\lambda4686$/H$\alpha$, He\,\textsc{i}$\lambda5876$/H$\alpha$,
He\,\textsc{ii}$\lambda4686$/H$\beta$,
He\,\textsc{i}$\lambda5876$/H$\beta$,
He\,\textsc{i}$\lambda5876$/He\,\textsc{ii}$\lambda4686$,
C\,\textsc{iv}/Ly$\alpha$,
He\,\textsc{ii}/Ly$\alpha$,
and Si\,\textsc{iv}/Mg\,\textsc{ii}. For optical emission lines, H$\alpha$ was used as the normalization line when available.
Otherwise, ratios relative to $\mathrm{H}\beta$ were adopted. These emission-line ratios were selected as screening criteria for constraining $\mathit{\Phi}_{\mathrm{out}}$ and $s$. 

We set a wide range of parameters: the outer boundary of the ionizing photon flux $\log_{10}(\mathit{\Phi}_{\mathrm{out}}/(\mathrm{cm^{-2}\,s^{-1}})) \in [17, 20]$, and the power-law index of the cloud distribution $s \in [-3, 0]$. To determine the optimal parameters for each source, we minimized the differences between the LOC-predicted and observationally measured emission-line ratios. The observed line ratios, $R_{\mathrm{obs}, i}$, and their corresponding uncertainties, $\sigma_{\mathrm{obs}, i}$, were derived from the emission-line fluxes. The flux uncertainties were estimated via Monte Carlo resampling of the spectra based on the spectral noise (as implemented in \texttt{PyQSOFit}), with 200 realizations per source. The uncertainty of each line ratio was then calculated through standard error propagation.

For each emission-line ratio $i$, we calculated the weighted difference between the observation and the model prediction $R_{\mathrm{model}, i}(\mathit{\Phi}_{\mathrm{out}}, s)$:
\begin{equation}
\chi^2_i = \left( \frac{R_{\mathrm{obs}, i} - R_{\mathrm{model}, i}(\mathit{\Phi}_{\mathrm{out}}, s)}{\sigma_{\mathrm{obs}, i}} \right)^2.
\end{equation}
Due to the inherent limitations of \texttt{Cloudy} \citep{ferland2017}, not all line ratios can be accurately reproduced. We therefore applied a two-step clipping procedure to reject poorly fitted line ratios. First, we discard any line ratio whose minimum $\chi^2_i$ value across the parameter grid (denoted as $\chi^2_{\min, i}$) exceeded 10. Second, we rejected line ratios that act as statistical outliers, defined as those whose $\chi^2_{\min, i}$ fall outside $\pm 2\,\mathrm{NMAD}$ of the median of the $\chi^2_{\min}$ distribution. The normalized median absolute deviation (NMAD) is calculated as:
\begin{equation}
\mathrm{NMAD} = 1.4826 \times \mathrm{median} \left( \left| \chi^2_{\min, i} - \mathrm{median}(\chi^2_{\min}) \right| \right),
\end{equation}
where the constant $1.4826$ ensures consistency with the standard deviation for a Gaussian distribution. These rejected line ratios are excluded from the subsequent determination of the optimal parameters.

After removing the outliers, we quantified the overall agreement using a mean $\chi^2$ statistic defined as:
\begin{equation}
\langle \chi^2 \rangle = \frac{1}{N} \sum_{i=1}^{N} \chi^2_i,
\end{equation}
where $N$ is the number of remaining valid emission-line ratios for a given source. The optimal parameters $(\mathit{\Phi}_{\mathrm{out}}, s)$ for each source are determined by minimizing $\langle \chi^2 \rangle$. To ensure robust parameter constraints and avoid degeneracy, we required at least two independent line ratios for cross-validation. Sources failing this criterion are excluded from our final sample; in this step, seven sources were removed (including Mrk~1018, which lacked significant broad emission lines, and six others with only one valid line ratio).

\begin{figure}
\includegraphics[width=1.\linewidth]{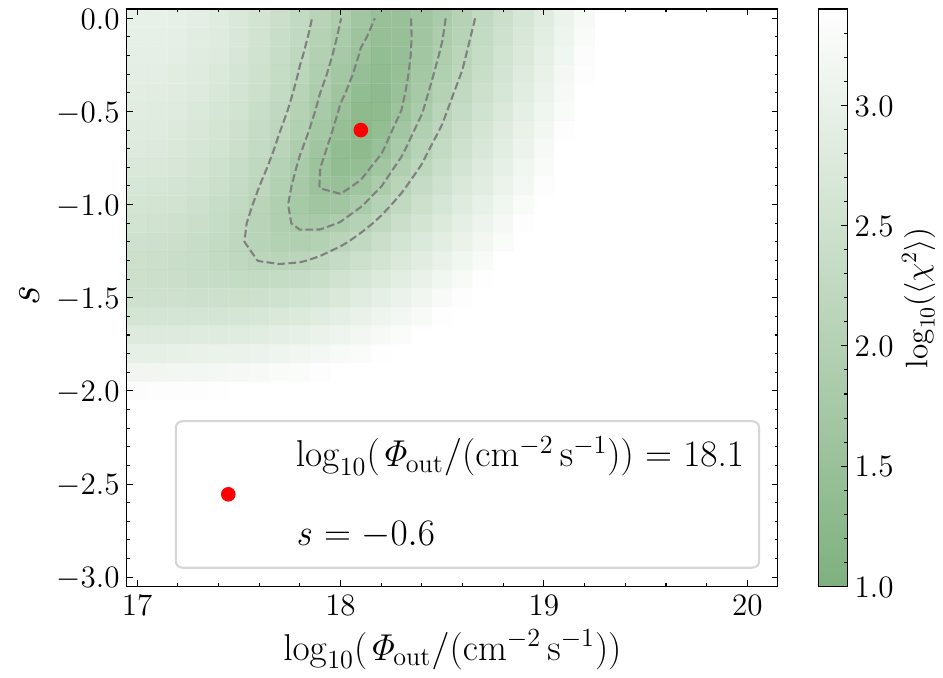}
\caption{$\langle\chi^{2}\rangle$ values for Mrk 509. The red points mark the parameter set corresponding to the minimum $\langle\chi^{2}\rangle$, which is adopted to construct the DC templates used in subsequent analysis. The gray dashed contours indicate levels at 2, 4, and 8 times the minimum $\langle\chi^{2}\rangle$ value.}
\label{Mrk279_line}
\end{figure}

Figure~\ref{Mrk279_line} presents an example of the optimal parameter determination result for Mrk~509. The adopted lines are listed in Table~\ref{tab:result}. The red point indicates the best-fitting parameters that minimize $\langle\chi^{2}\rangle$, namely $\mathit{s} = -0.6$ and $\mathrm{log}_{10}(\mathit{\Phi}_{\mathrm{out}}/\mathrm{cm^{-2}s^{-1})} = 18.1$, corresponding to an outer BLR radius of 323.19~lt-days. Using these optimal parameters, we constructed source-specific DC spectral templates by weighting the cloud distribution accordingly. These templates were subsequently adopted in the spectral decomposition analysis.

\begin{figure}
\includegraphics[width=1.\linewidth]{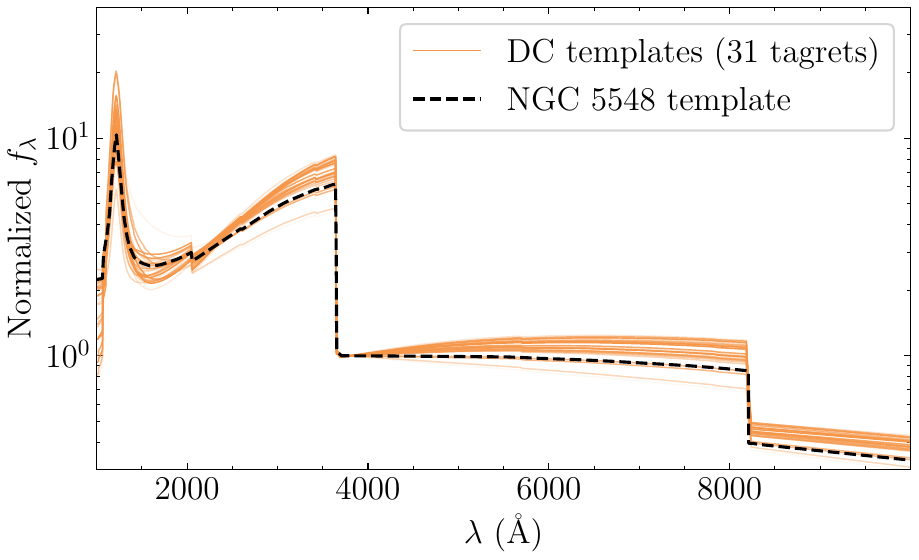}
\caption{Normalized DC templates for the 31} sources in our sample. Each orange curve represents an individual DC template from our sample. The black dashed curve shows the DC template of NGC 5548 from \cite{Korista2019}. All DC templates are normalized at rest-frame $3800\ \mathrm{\AA}$ to better highlight the differences in overall shape. Overall, the DC templates for the 31 sources resemble the features for NGC 5548, but the exact shape varies per source.
\label{DC_template_result}
\end{figure}

\section{Results and Discussion} \label{sec:res}

\subsection{DC template and decomposition results} \label{sec:decomp_results}

\figsetstart
\figsetnum{4}
\figsettitle{Spectral Fitting Results for 31 Sources}
\figsetgrpstart
\figsetgrpnum{4.1}
\figsetgrptitle{3C 382}
\figsetplot{3C382_result.pdf}
\figsetgrpnote{Spectral fitting result for 3C 382.}
\figsetgrpend

\figsetgrpstart
\figsetgrpnum{4.2}
\figsetgrptitle{Arp 151}
\figsetplot{Arp151_result.pdf}
\figsetgrpnote{Spectral fitting result for Arp 151.}
\figsetgrpend

\figsetgrpstart
\figsetgrpnum{4.3}
\figsetgrptitle{Mrk 50}
\figsetplot{Mrk50_result.pdf}
\figsetgrpnote{Spectral fitting result for Mrk 50.}
\figsetgrpend

\figsetgrpstart
\figsetgrpnum{4.4}
\figsetgrptitle{Mrk 279}
\figsetplot{Mrk279_result.pdf}
\figsetgrpnote{Spectral fitting result for Mrk 279.}
\figsetgrpend

\figsetgrpstart
\figsetgrpnum{4.5}
\figsetgrptitle{Mrk 817}
\figsetplot{Mrk817_result.pdf}
\figsetgrpnote{Spectral fitting result for Mrk 817.}
\figsetgrpend

\figsetgrpstart
\figsetgrpnum{4.6}
\figsetgrptitle{Mrk 1310}
\figsetplot{Mrk1310_result.pdf}
\figsetgrpnote{Spectral fitting result for Mrk 1310.}
\figsetgrpend

\figsetgrpstart
\figsetgrpnum{4.7}
\figsetgrptitle{NGC 6814}
\figsetplot{NGC6814_result.pdf}
\figsetgrpnote{Spectral fitting result for NGC 6814.}
\figsetgrpend

\figsetgrpstart
\figsetgrpnum{4.8}
\figsetgrptitle{PG 1226+023}
\figsetplot{PG1226+023_result.pdf}
\figsetgrpnote{Spectral fitting result for PG 1226+023.}
\figsetgrpend

\figsetgrpstart
\figsetgrpnum{4.9}
\figsetgrptitle{PHL 1811}
\figsetplot{PHL1811_result.pdf}
\figsetgrpnote{Spectral fitting result for PHL 1811.}
\figsetgrpend

\figsetgrpstart
\figsetgrpnum{4.10}
\figsetgrptitle{SBS 1116+583A}
\figsetplot{SBS1116+583A_result.pdf}
\figsetgrpnote{Spectral fitting result for SBS 1116+583A.}
\figsetgrpend

\figsetgrpstart
\figsetgrpnum{4.11}
\figsetgrptitle{ZW 229-015}
\figsetplot{ZW229-015_result.pdf}
\figsetgrpnote{Spectral fitting result for ZW 229-015.}
\figsetgrpend

\figsetgrpstart
\figsetgrpnum{4.12}
\figsetgrptitle{Mrk 110}
\figsetplot{Mrk110_result.pdf}
\figsetgrpnote{Spectral fitting result for Mrk 110.}
\figsetgrpend

\figsetgrpstart
\figsetgrpnum{4.13}
\figsetgrptitle{Mrk 493}
\figsetplot{Mrk493_result.pdf}
\figsetgrpnote{Spectral fitting result for Mrk 493.}
\figsetgrpend

\figsetgrpstart
\figsetgrpnum{4.14}
\figsetgrptitle{Mrk 509}
\figsetplot{Mrk509_result.pdf}
\figsetgrpnote{Spectral fitting result for Mrk 509.}
\figsetgrpend

\figsetgrpstart
\figsetgrpnum{4.15}
\figsetgrptitle{NGC 4151}
\figsetplot{NGC4151_result.pdf}
\figsetgrpnote{Spectral fitting result for NGC 4151.}
\figsetgrpend

\figsetgrpstart
\figsetgrpnum{4.16}
\figsetgrptitle{NGC 4593}
\figsetplot{NGC4593_result.pdf}
\figsetgrpnote{Spectral fitting result for NGC 4593.}
\figsetgrpend

\figsetgrpstart
\figsetgrpnum{4.17}
\figsetgrptitle{ESO 362-G18}
\figsetplot{ESO362-G18_result.pdf}
\figsetgrpnote{Spectral fitting result for ESO 362-G18.}
\figsetgrpend

\figsetgrpstart
\figsetgrpnum{4.18}
\figsetgrptitle{ESO 511-30}
\figsetplot{ESO511-30_result.pdf}
\figsetgrpnote{Spectral fitting result for ESO 511-30.}
\figsetgrpend

\figsetgrpstart
\figsetgrpnum{4.19}
\figsetgrptitle{Fairall 9}
\figsetplot{Fairall9_result.pdf}
\figsetgrpnote{Spectral fitting result for Fairall 9.}
\figsetgrpend

\figsetgrpstart
\figsetgrpnum{4.20}
\figsetgrptitle{H 0507+164}
\figsetplot{H0507+164_result.pdf}
\figsetgrpnote{Spectral fitting result for H 0507+164.}
\figsetgrpend

\figsetgrpstart
\figsetgrpnum{4.21}
\figsetgrptitle{Mrk 590}
\figsetplot{Mrk590_result.pdf}
\figsetgrpnote{Spectral fitting result for Mrk 590.}
\figsetgrpend

\figsetgrpstart
\figsetgrpnum{4.22}
\figsetgrptitle{Mrk 1044}
\figsetplot{Mrk1044_result.pdf}
\figsetgrpnote{Spectral fitting result for Mrk 1044.}
\figsetgrpend

\figsetgrpstart
\figsetgrpnum{4.23}
\figsetgrptitle{Mrk 1048}
\figsetplot{Mrk1048_result.pdf}
\figsetgrpnote{Spectral fitting result for Mrk 1048.}
\figsetgrpend

\figsetgrpstart
\figsetgrpnum{4.24}
\figsetgrptitle{Mrk 1239}
\figsetplot{Mrk1239_result.pdf}
\figsetgrpnote{Spectral fitting result for Mrk 1239.}
\figsetgrpend

\figsetgrpstart
\figsetgrpnum{4.25}
\figsetgrptitle{Mrk 1298}
\figsetplot{Mrk1298_result.pdf}
\figsetgrpnote{Spectral fitting result for Mrk 1298.}
\figsetgrpend

\figsetgrpstart
\figsetgrpnum{4.26}
\figsetgrptitle{NGC 2617}
\figsetplot{NGC2617_result.pdf}
\figsetgrpnote{Spectral fitting result for NGC 2617.}
\figsetgrpend

\figsetgrpstart
\figsetgrpnum{4.27}
\figsetgrptitle{NGC 3783}
\figsetplot{NGC3783_result.pdf}
\figsetgrpnote{Spectral fitting result for NGC 3783.}
\figsetgrpend

\figsetgrpstart
\figsetgrpnum{4.28}
\figsetgrptitle{NGC 4748}
\figsetplot{NGC4748_result.pdf}
\figsetgrpnote{Spectral fitting result for NGC 4748.}
\figsetgrpend

\figsetgrpstart
\figsetgrpnum{4.29}
\figsetgrptitle{PG 2214+139}
\figsetplot{PG2214+139_result.pdf}
\figsetgrpnote{Spectral fitting result for PG 2214+139.}
\figsetgrpend

\figsetgrpstart
\figsetgrpnum{4.30}
\figsetgrptitle{PG 0934+013}
\figsetplot{PG0934+013_result.pdf}
\figsetgrpnote{Spectral fitting result for PG 0934+013.}
\figsetgrpend

\figsetgrpstart
\figsetgrpnum{4.31}
\figsetgrptitle{PKS 2004-447}
\figsetplot{PKS2004-447_result.pdf}
\figsetgrpnote{Spectral fitting result for PKS 2004-447.}
\figsetgrpend

\figsetend

\begin{figure*}
\includegraphics[width=\textwidth]{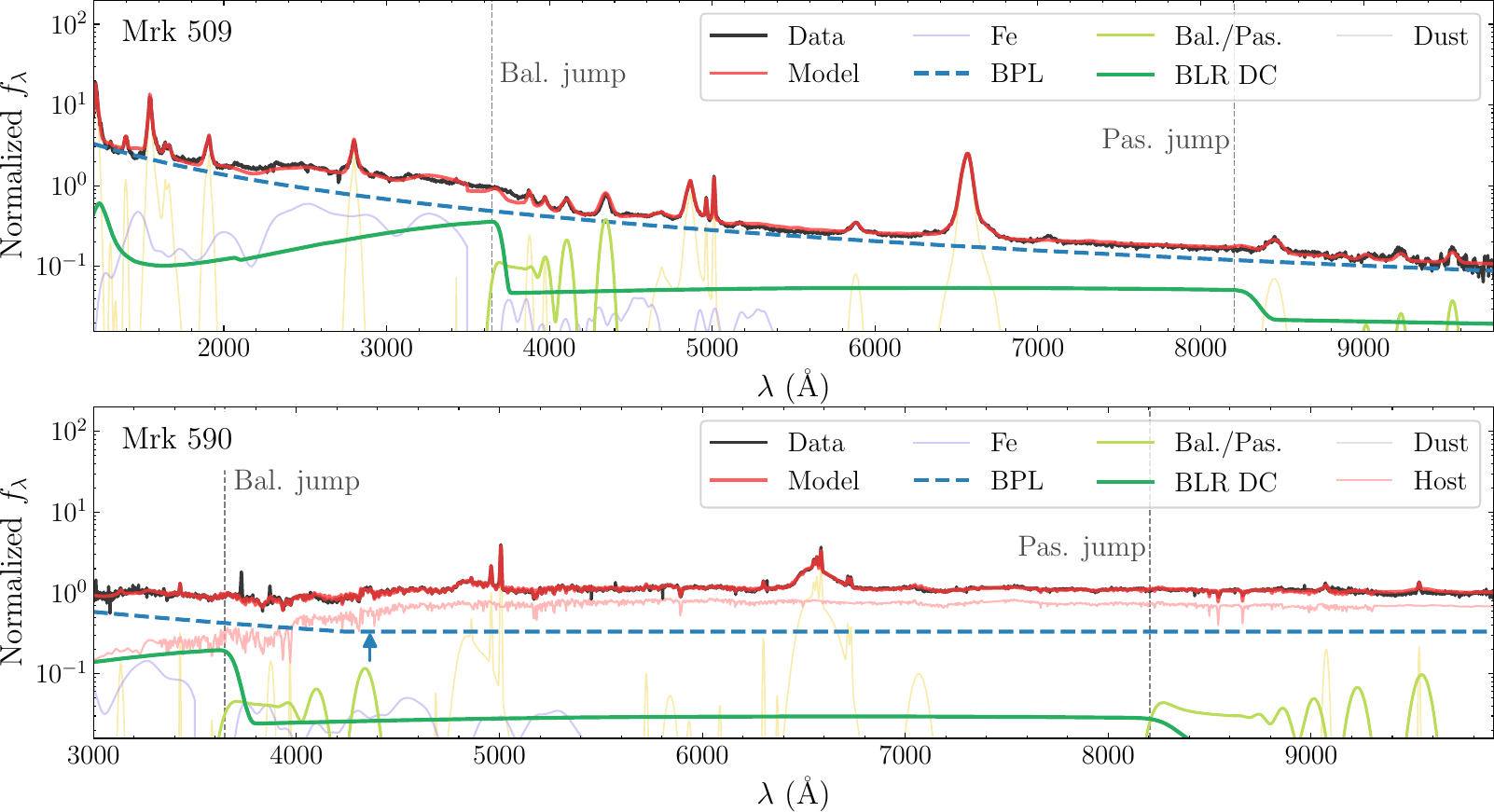}
\caption{Spectral decomposition results for Mrk~509 and Mrk~590}. Each spectrum is normalized to its median flux. The spectra for Mrk~509 and Mrk~590 are obtained with \textit{HST}/STIS and VLT/X-shooter, respectively. In each panel, the black and red curves show the observed spectrum and the best-fit model, respectively. Individual model components include the accretion-disk emission (blue dashed line), the DC component (dark green curve), the high-order Balmer and Paschen line series (light green curve), the host-galaxy (pink curve), the Fe\,{\sc ii} emission complexes (purple curve), and other emission lines (yellow curve). The blue arrows indicate the break wavelengths of the BPL model. Panels without a blue arrow correspond to cases where the fitted accretion-disk continuum is a single power law (PL). The gray dashed vertical lines mark the wavelengths of the Balmer jump at 3646 \AA~ and the Paschen jump at 8206 \AA. The complete figure set (31 images) is available in the online journal.
\label{decomposition_result}
\end{figure*}

\begin{figure}
\includegraphics[width=1.\linewidth]{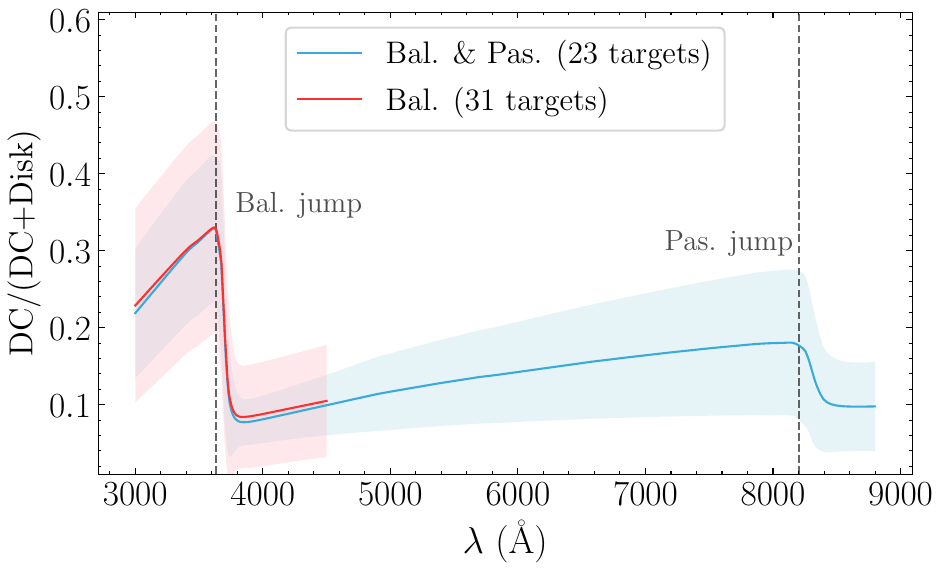}
\caption{Average DC fraction, derived from spectral fitting. The blue curve shows the results for sources whose spectra simultaneously cover both the Balmer and Paschen jumps, while the pink curve includes all 31} sources in our sample, including those covering only the Balmer jump. In both cases, the shaded regions indicate the corresponding $1\sigma$ standard deviations. The results demonstrate that the DC component makes a substantial contribution to the AGN continuum.
\label{DC_fraction}
\end{figure}

Our final sample has 31 AGNs with redshifts ranging from 0.003 to 0.24, and a median redshift of 0.027. We obtained 23 spectra covering both the Balmer and Paschen jumps (nine from the \textit{HST}/STIS and 14 from VLT/X-shooter), along with eight spectra covering only the Balmer jump (seven from the \textit{HST}/STIS and one from VLT/X-shooter). Table \ref{tab:obs_t} summarizes the observation information and the AGN parameters, and Table \ref{tab:result} lists the BLR parameters used for the DC template generation and the DC contribution from the spectral fitting. Furthermore, we compared the BLR outer radius ($R_{\mathrm{out}}$, derived from the best-fit value of $\mathit{\Phi}_{\mathrm{out}}$) with the H$\beta$ radius ($R_{\mathrm{H}\beta}$, obtained from reverberation mapping). The ratio of these two radii yields a mean $\log_{10}(R_{\mathrm{out}}/R_{\mathrm{H}\beta}) = 1.06 \pm 0.56$, which is roughly consistent with the dust reverberation mapping results \citep[e.g.,][]{Lyu2019}. 

\begin{figure*}
\includegraphics[width=1.\linewidth]{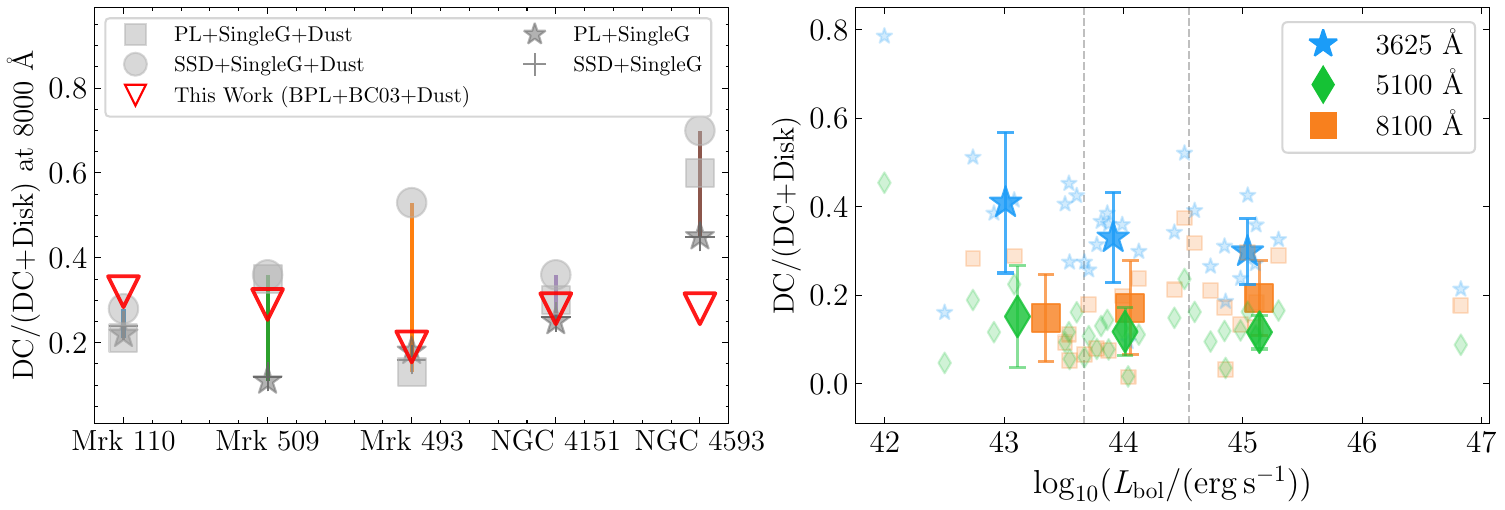}
\caption{\textit{Left:} Comparison of our best-fit results with their various results reported by \cite{Guo_2022} at 8000 \AA. The gray points represent their estimates, while the red inverted triangles denote our derived values. ``PL'', ``SSD'', and ``SingleG'' refer to the single power-law, standard disk, and single-age stellar population host models, respectively. Most of our results are generally consistent with the range spanned by their various model configurations.} \textit{Right:} The dependence of the DC fraction on bolometric luminosity. The small blue stars, green diamonds, and orange squares represent the DC contributions at 3625~\AA, 5100~\AA, and 8100~\AA, respectively, for all sources except PHL~1811. The 30 sources are divided into three bolometric-luminosity bins, with the vertical dashed lines indicating the bin boundaries. The large blue stars, green diamonds, and orange squares, together with their corresponding error bars, indicate the mean DC contributions and $1\sigma$ standard deviations. The green diamonds are plotted at the mean bolometric luminosity of each bin, while the blue stars and orange squares are offset by $-0.1$ and $+0.1$ dex in the x-axis, respectively, for clarity. The DC fraction near the Balmer jump seems to decrease with increasing bolometric luminosity.
\label{DC_compare}
\end{figure*}

Figure~\ref{DC_template_result} shows the DC templates normalized to the luminosity at 3800~\AA. The resulting templates exhibit Balmer (3646~\AA) and Paschen (8206~\AA) jump features, which well resemble NGC~5548 presented by \cite{Korista2019}. However, the strengths of these jumps vary among sources. In addition, the continuum slopes differ significantly both in the optical region between the two jumps and at wavelengths shortward of the Balmer jump. These variations likely reflect differences in the BLR cloud distribution and the shape of the ionizing SED. The diversity in continuum slopes across the optical bands underscores the importance of constructing source-specific DC templates.

We implemented the source-specific DC templates in the spectral decomposition code \texttt{PyQSOFit}. Then, we used this code to fit the \textit{HST} or VLT spectra of our sources. In this step, for each of the 31 sources, we used its specific DC template shape and allowed the fraction of DC emission to vary freely. Here, we present two spectral decomposition results for Mrk~509, and Mrk~590 in Figure~\ref{decomposition_result}. The spectra for Mrk~509 and Mrk~590 are obtained with \textit{HST}/STIS and VLT/X-shooter, respectively.

 We determined the best-fitting DC fraction for each source, defined as $f_\mathrm{DC}/(f_\mathrm{DC} + f_\mathrm{disk})$. Figure~\ref{DC_fraction} displays the average DC fraction obtained from spectral fitting. We quantified the DC contribution using DC fractions at 3625~\AA~(before the Balmer jump), 5100~\AA, and 8100~\AA~(before the Paschen jump) to describe the contribution at different wavelengths. The DC fractions for each source are listed in Table~\ref{tab:result}. For the 23 sources whose rest-frame spectra cover both the Balmer and Paschen jumps, the average DC fraction is \(0.33 \pm 0.09\) at~3625~\AA, \(0.12 \pm 0.05\) at 5100~\AA, and \(0.18 \pm 0.09\) at 8100~\AA. For the 31 sources whose rest-frame spectra at least cover the Balmer jump, the mean DC fraction is $0.33 \pm 0.14$ at 3625~\AA. At 5100~\AA, the mean DC fraction is $0.13 \pm 0.08$ for the remaining 30 sources after excluding PHL~1811, whose rest-frame spectral coverage does not extend to 5100~\AA. Among the 31 sources, Mrk~1239 and PHL~1811 have particularly low DC contributions. Hence, we expect the continuum reverberation mapping of Mrk~1239 to test the disk size, as its optical variability is quite obvious. 
 
To investigate whether host-galaxy contamination could introduce a systematic bias in the inferred DC fractions, we compare the mean DC fraction at 3625~\AA\ between HST/STIS sources with negligible host-galaxy contributions that cover both Balmer and Paschen jumps and the four sources with significant host-galaxy contributions. The mean DC fractions for these two groups are \(0.38 \pm 0.10\) and \(0.34 \pm 0.04\), respectively. We find no obvious difference between the two groups, suggesting that host-galaxy contamination does not appear to strongly affect the inferred DC fractions in our sample.

\begin{figure*} 
\includegraphics[width=1.\linewidth]{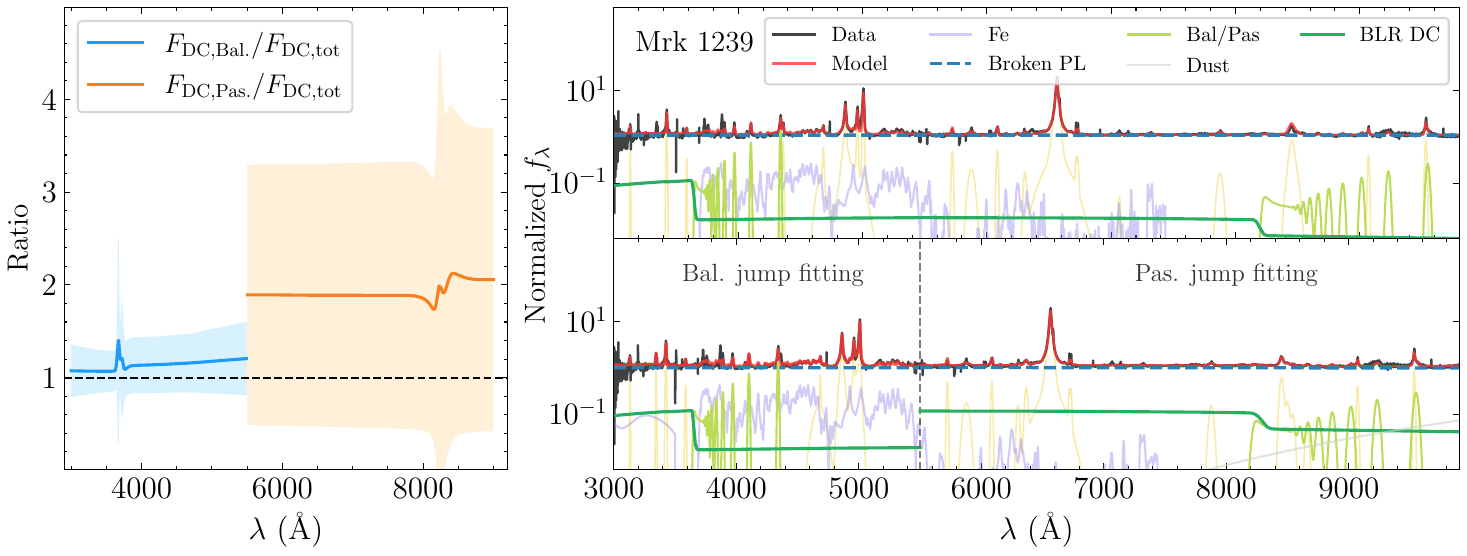}
\caption {\textit{Left:} Comparison of the DC components derived from single-jump fitting under limited spectral coverage with those from two-jump fitting. Shown are the ratios of the DC fractions derived using only the Balmer jump (blue) and only the Paschen jump (orange), $F_{\mathrm{DC, part}}$, relative to those derived from fitting both jumps, $F_{\mathrm{DC, tot}}$. The shaded regions show the 1$\sigma$ dispersion. The results indicate that Balmer-jump fits are more robust, while Paschen-jump fits are usually overestimated. \textit{Right:} Example of the spectral decomposition results of Mrk~1239, comparing the results obtained from two-jump fitting and single-jump fitting. The top panel shows the decomposition result obtained from fitting both jumps, while the bottom panel shows the decomposition results obtained by fitting only the Balmer jump and only the Paschen jump. The gray dashed line marks the wavelength to separate the blue and red spectral regions.}
\label{Single_jump}
\end{figure*}

The gap of the Fe\,{\sc ii} emission template around the Balmer jump (see the purple curves in Figure~\ref{decomposition_result}) may lead to a modest overestimation of the DC contribution. To verify this effect, we performed a simple linear interpolation between the UV and optical Fe\,{\sc ii} emission templates to fill this gap; the gap-filled iron template is used to refit the spectra of NGC~6814 and Mrk~493, both of which exhibit strong iron emission, and Mrk~1048, which shows relatively weak Fe\,{\sc ii} emission. The results indicate that the DC contribution is reduced by approximately $5\%-15\%$ compared with the original decomposition. Despite these uncertainties, our results demonstrate that the DC component makes a substantial contribution to the AGN continuum, especially around the two jumps.

We further compared our results with those reported by \citet{Guo_2022}, who estimated the DC contribution in the near-infrared band (6800--9700~\AA) focusing on the Paschen jump. They also tested their results by changing the disk emission prescription from the broken power law to either a single power law or the standard thin disk model of \citet{Shakura_1973}. In addition, they used a single stellar population model to describe the host galaxy component. The comparison results for the five sources are presented in the left panel of Figure~\ref{DC_compare}. For Mrk~493 and NGC~4151, our derived DC contributions lie near the lower end of the range reported by \citet{Guo_2022}, whereas those for Mrk~110 and Mrk~509 are closer to the upper end. In contrast, our estimate for NGC~4593 is substantially lower than their reported range. Overall, most of our estimates are consistent with the range of DC contributions yielded by their analyses.

\subsection{DC fraction with bolometric luminosity}
We further investigated the dependence of the DC fraction on bolometric luminosity. Based on the decomposition results, we calculated the absolute flux of the decomposed disk components at 5100~\AA\ and estimated the corresponding bolometric luminosity using $L_{\text{bol}} = 10\,L_{5100\,\text{\AA}}$ \citep{Richards_2006}. The rest-frame wavelengths of PHL~1811 do not extend to 5100~\AA, and it was not included in this analysis. The remaining 30 sources were divided into three luminosity bins according to $L_{\mathrm{bol}}$, each containing 10 sources, respectively. Some spectra do not cover 8100~\AA, the corresponding sample sizes for 8100~\AA\ in each bin are 6, 8, and 10. For each bin, we calculated the mean DC fraction and its standard deviation at 3625~\AA, 5100~\AA, and 8100~\AA. The results are shown in the right panel of Figure~\ref{DC_compare}. 

Our results show that the DC contribution around the Balmer jump is likely luminosity-dependent, with lower-luminosity sources exhibiting higher DC fractions. In contrast, at wavelengths near the Paschen jump and 5100~\AA, the three luminosity bins roughly share the similar DC fractions. The DC fraction--luminosity relation in the Balmer jump is qualitatively consistent with the Baldwin effect \citep{Baldwin_effect1977} observed in broad emission lines (e.g., C\,{\sc iv}). However, \citet{Pan_2026} reported that the DC component accounts for only $\sim$2.7\% of the total flux around the Balmer jump in the low-luminosity intermediate-mass black hole NGC~4395 ($L_{\rm bol} \sim 1.52 \times 10^{41}$ erg s$^{-1}$), and their high-cadence continuum reverberation mapping also detected no significant $u$-band lag excess. Regardless of these comparisons, given the limited sample size and the substantial dispersion within each luminosity bin, we therefore refrain from drawing firm conclusions regarding a luminosity dependence of the DC contribution, which requires a larger and more comprehensive sample size for further validation.

\subsection{DC fraction with wavelength coverage}

Due to the lack of sufficiently wide wavelength coverage for many AGNs, one may only use either the Balmer or Paschen jump to perform the spectral decomposition. In this case, biases might be introduced. 

To simulate the case in which only one discontinuity is covered, we focused on the 23 sources whose spectra cover both the Balmer and Paschen jumps. For sources with a host galaxy contribution, we first subtracted this component. We then separately decomposed the blue ($\lambda<5500\,\mathrm{\AA}$) and red ($\lambda>5500\,\mathrm{\AA}$) halves of the spectrum for each of the 23 sources. During the spectral decomposition, we tied the DC velocity broadening to that of the corresponding high-order emission lines (the Balmer series for the blue half and the Paschen series for the red half) to constrain the velocity of the DC component. For each source, we computed the ratio between the DC fraction derived from the Balmer jump alone (hereafter, $F_{\mathrm{DC,\,Bal.}}$) and that obtained from both jumps, $F_{\mathrm{DC,\,tot}}$. We performed a similar calculation for the DC fraction derived from the Paschen jump alone (hereafter, $F_{\mathrm{DC,\,Pas.}}$). 

The left panel of Figure~\ref{Single_jump} presents the mean values and $1\sigma$ standard deviation of the ratio $F_{\mathrm{DC,Bal.}}/F_{\mathrm{DC,tot}}$ and $F_{\mathrm{DC,Pas.}}/F_{\mathrm{DC,tot}}$ for the 23 sources. The ratio $F_{\mathrm{DC,Bal.}}/F_{\mathrm{DC, tot}}$ remains close to 1, suggesting that the Balmer jump alone provides a relatively reliable constraint on the DC component. In contrast, $F_{\mathrm{DC, Pas.}}/F_{\mathrm{DC, tot}}$ exhibits a mean value of approximately 2 with substantial dispersion, implying that fits restricted to the Paschen region overestimate the DC fraction by a factor of two on average. The right panel of Figure~\ref{Single_jump} shows the single-jump decomposition for Mrk~1239, which represents the most extreme case in our sample. When only the Paschen jump is used, the inferred DC fraction is overestimated by a factor of $\sim$6.5. This source also has one of the weakest intrinsic DC contributions in the sample.

To further investigate the origin of this overestimation, we split the 23 sources into two bins at the median $F_{\mathrm{DC, Pas.}}/F_{\mathrm{DC,\,tot}}$ at 8100$\,\mathrm{\AA}$ of 1.26. We find that sources with severe overestimation have an average intrinsic DC fraction $F_{\mathrm{DC,\,tot}}(8100\,\mathrm{\AA})$ of only $\sim$0.12, whereas the remaining sources show a higher average of $\sim$0.23. This indicates that overestimation primarily occurs when the intrinsic DC contribution is weak. In such cases, the intrinsically weak Paschen jump is insufficient to constrain the DC fraction. Overall, these results demonstrate that spectra with Balmer jump coverage are required to robustly constrain the DC contribution.

\subsection{Error analysis}
\begin{figure} 
\includegraphics[width=1.\linewidth]{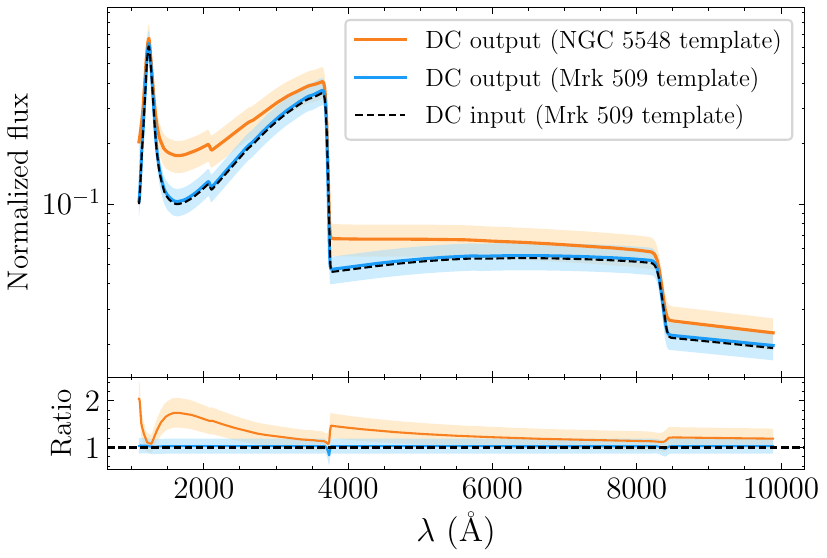}
\caption {\textit{Left:} The results of DC component fitting for 100 simulated spectra with randomly perturbed accretion-disk components. The top panel presents the true DC component (black dashed line), and the median recovered DC components with different templates (solid curves). Shaded regions indicate the NMAD. Blue and orange curves indicate results from Mrk~509 and NGC~5548, respectively. The bottom panel gives the ratio of recovered to true DC components. These results demonstrate that the DC component can be recovered with a suitable template. An inappropriate template introduces wavelength-dependent systematic biases.}
\label{DC_both}
\end{figure}

We use Mrk~509 to demonstrate the uncertainties in the spectral decomposition. We used the best-fit decomposed components to generate mock spectra by varying the relative contribution of the accretion disk component. Specifically, we scaled the original disk fraction, $f_{\mathrm{disk}}$, by a factor $k$, while keeping the other components fixed.

To explore a broad range of disk contributions, we randomly drew 100 values $n_i$ from the interval $[-1,1]$ and adopted $k = 10^{n_i}$ as the scaling factor. Hence, the disk contribution in the mock spectra can be 10 times larger or smaller than the best-fitting one. We incorporated observational uncertainties into the simulated spectra by introducing a white-noise term based on the flux errors of the original spectrum. Specifically, we constructed mock noisy data as 
\begin{equation}
f_\mathrm{mock} = f_\mathrm{tot} * (1 + \delta) ,
\end{equation}
where $\delta$ is a random variable drawn from a normal distribution with mean zero and standard deviation $ \sigma$ ($\sigma = \mathrm{err}/\mathrm{data}$) evaluated at each wavelength point. We generated 100 mock spectra and decomposed each using our fitting procedure, thereby obtaining the corresponding DC components. In the spectral decomposition of the mock spectra, we always adopted our LOC calculation of the DC template for Mrk~509.

Figure~\ref{DC_both} presents the median and NMAD of the recovered DC components. When fitting with the Mrk~509 template, the recovered-to-input DC ratio is \( 1.03 \pm 0.16 \), excluding small fluctuations near the Balmer and Paschen jumps that arise from the allowed 5-pixel wavelength shift during the fitting process. This result indicates that our decomposition procedure reliably recovers the input DC component when the correct template is adopted. 

In this test, we also try to use the DC template for NGC~5548 (rather than for Mrk~509) in the spectral decomposition analysis. This test was designed to quantify how the choice of DC template influences the recovered DC component in spectral decomposition and to assess the resulting impact on corrections to the inferred accretion disk emission. Compared with the Mrk~509 DC template, fitting with the NGC~5548 DC template yields systematically different DC fractions across wavelengths. Here, we report the ratios of recovered-to-input ratios with the NGC~5548 DC template at three representative wavelengths: 3500~\AA\ and 3800~\AA, corresponding to the pre- and post-Balmer jump regions, respectively, and 8100~\AA\ near the Paschen jump. The recovered-to-input ratios are \( 1.16 \pm 0.21 \) at 3500~\AA, \(1.46 \pm 0.27\) at 3800~\AA, and \(1.14 \pm 0.21\) at 8100~\AA. These values indicate that adopting the NGC~5548 template generally leads to an overestimation of the DC contribution across different wavelength regimes for the Mrk~509 spectrum. This overestimation would enlarge the continuum time delays predicted by the SSD model with the DC contribution for Mrk~509.

\begin{deluxetable*}{llcccccc}
\tablenum{2}
\tablecaption{Sample template parameters and DC fraction \label{tab:result}}
\tabletypesize{\footnotesize}
\tablehead{
\colhead{Object} & \colhead{Emission Lines} & \colhead{$R_{\mathrm{in}}$} & \colhead{$\log_{10} \mathit{\Phi}_{\mathrm{out}}$} & \colhead{$s$} & \multicolumn{3}{c}{Ratio} \\
\cline{6-8}
\colhead{} & \colhead{} & \colhead{(lt-days)} & \colhead{$\log_{10} \mathrm{(cm^{-2}\,s^{-1}})$} & \colhead{} & \colhead{3625~\AA} & \colhead{5100~\AA} & \colhead{8100~\AA}
}
\colnumbers{}
\startdata
ESO 362-G18&  H$\beta$/H$\alpha$, He\,{\sc ii}~4686/H$\alpha$, He\,{\sc i}~5876/H$\alpha$
&1.08 &	18.3 &	-0.1 & 40\% &9\% &9\%\\
ESO 511-30&		H$\beta$/H$\alpha$, He\,{\sc ii}~4686/H$\alpha$	&1.87&17.8	& -0.8
&36\%&8\%&8\%\\
Fairall 9 &		H$\beta$/H$\alpha$, He\,{\sc i}~5876/H$\alpha$	&1.74&17.1	&-1.6	
&24\%&12\%&13\%\\
H 0507+164&		H$\beta$/H$\alpha$, He\,{\sc ii}~4686/H$\alpha$, He\,{\sc i}~5876/H$\alpha$	&0.30&17.0	&-0.2&43\%&12\%&11\%				\\	
Mrk 590	&H$\beta$/H$\alpha$, He\,{\sc i}~5876/H$\alpha$&	7.96&17.9&	-0.2	
&31\%&8\%&8\%\\		
Mrk 1044	&H$\beta$/H$\alpha$, He\,{\sc ii}~4686/H$\alpha$&	1.05&19.1	&0	&29\%&11\%&24\%						
\\
Mrk 1048	&H$\beta$/H$\alpha$, He\,{\sc ii}~4686/H$\alpha$, He\,{\sc i}~5876/H$\alpha$&0.74&	18.1&	-0.3	
&31\%&12\%&17\%\\
Mrk 1239&H$\beta$/H$\alpha$, He\,{\sc ii}~4686/H$\alpha$, He\,{\sc i}~5876/H$\alpha$&1.94&	17.2	&-1.1&	9\%&2\%&1\%\\				
Mrk 1298	&H$\beta$/H$\alpha$, He\,{\sc ii}~4686/H$\alpha$, He\,{\sc i}~5876/H$\alpha$	&16.00&18.5	&0&35\%&11\%&18\%	\\			
NGC 2617	&H$\beta$/H$\alpha$, He\,{\sc i}~5876/H$\alpha$&	0.54&17.0&	-1.1	
&27\%&6\%&5\%\\
NGC 3783&H$\beta$/H$\alpha$, He\,{\sc ii}~4686/H$\alpha$, He\,{\sc i}~5876/H$\alpha$ &0.95&	18.3	&-0.2
&25\%&11\%&18\%\\
NGC 4748	&H$\beta$/H$\alpha$, He\,{\sc ii}~4686/H$\alpha$, He\,{\sc i}~5876/H$\alpha$&0.56&18.4	&-0.8&	26\%&6\%&6\%			\\	
PG 2214+139	&H$\beta$/H$\alpha$, He\,{\sc i}~5876/H$\alpha$&2.80&	17.0	&-1.8	
&27\%&11\%&13\%\\
PG 0934+013	&H$\beta$/H$\alpha$, He\,{\sc ii}~4686/H$\alpha$, He\,{\sc i}~5876/H$\alpha$& 0.81&	19.5	&0	 &33\%&15\%&21\%\\
PKS 2004-447&H$\beta$/H$\alpha$, He\,{\sc ii}~4686/H$\alpha$&	5.85&19.7	&0	&18\%&4\%&3\%					
\\
3C 382&	(H$\beta$+He\,{\sc ii}~4686)/H$\alpha$, He\,{\sc i}~5876/H$\alpha$	&4.04&18.6&	-0.1&	32\%&16\%&29\%						\\			
Arp 151 &C\,{\sc iv}/Ly$\alpha$, He\,{\sc ii}~1640/Ly$\alpha$	&11.3&17.0	&-1.4		
&38\%&12\%&-\\										
Mrk 50&	C\,{\sc iv}/Ly$\alpha$, He\,{\sc ii}~1640/Ly$\alpha$, Si\,\textsc{iv}/Mg\,\textsc{ii}	&1.04&18.4&	-0.3		&36\%&13\%&-				
\\							
Mrk 279&	H$\beta$/H$\alpha$, He\,{\sc i}~5876/H$\alpha$, C\,{\sc iv}/Ly$\alpha$,	&1.67&18.4&	-1.1&	52\%&24\%&37\%
\\		
&  	He\,{\sc ii}~1640/Ly$\alpha$, Si\,\textsc{iv}/Mg\,\textsc{ii}&	  &  
\\
Mrk 817	&		H$\beta$/H$\alpha$, C\,{\sc iv}/Ly$\alpha$, He\,{\sc ii}~1640/Ly$\alpha$,	&1.40&17.5&	0&	26\%&9\%&21\%		\\	
 &  Si\,\textsc{iv}/Mg\,\textsc{ii}&	  &  
\\
Mrk 1310&		C\,{\sc iv}/Ly$\alpha$, He\,{\sc ii}~1640/Ly$\alpha$, Si\,\textsc{iv}/Mg\,\textsc{ii}	&0.37&19.1&	-0.7			&16\%&5\%&-	\\								
NGC 6814	&	He\,{\sc ii}~1640/Ly$\alpha$, Si\,\textsc{iv}/Mg\,\textsc{ii}	&0.66&18.5&	-1.5			&78\%&45\%&-				
\\						
PG 1226+023	&	H$\beta$/H$\alpha$, He\,{\sc i}~5876/H$\alpha$&	30.68&18.3&	-1.8			&21\%&9\%&18\%
\\	
PHL 1811&	He\,{\sc ii}~1640/Ly$\alpha$, Si\,\textsc{iv}/Mg\,\textsc{ii}	&19.95&17.7	&0&3\%&-&-
\\
SBS 1116+583A	 &He\,{\sc ii}~4686/H$\beta$, C\,{\sc iv}/Ly$\alpha$, He\,{\sc ii}~1640/Ly$\alpha$,	&0.23&17.1	&-1.2&	42\%&16\%&-
\\
 &Si\,\textsc{iv}/Mg\,\textsc{ii}&	  & 
\\
ZW 229-015&		He\,{\sc ii}~4686/H$\beta$, C\,{\sc iv}/Ly$\alpha$, He\,{\sc ii}~1640/Ly$\alpha$,	&0.39&17.9&	-1.1&38\%&14\%&-\\							
&Si\,\textsc{iv}/Mg\,\textsc{ii}&	  & \\
Mrk 110	&		H$\beta$/H$\alpha$, He\,{\sc ii}~4686/H$\alpha$, He\,{\sc i}~5876/H$\alpha$	&2.78&19.4	&0	&	38\%&16\%&32\%	\\	 
Mrk 493	&		H$\beta$/H$\alpha$, He\,{\sc ii}~4686/H$\alpha$, C\,{\sc iv}/Ly$\alpha$,&1.16&17.0	&-0.9& 35\%&12\%&20\%\\
 &He\,{\sc ii}~1640/Ly$\alpha$, Si\,\textsc{iv}/Mg\,\textsc{ii}&	  & 
\\		
Mrk 509	&	H$\beta$/H$\alpha$, He\,{\sc ii}~4686/H$\alpha$, He\,{\sc i}~5876/H$\alpha$,	&7.96&18.1	&-0.6 &42\%&16\%&30\%\\
 & C\,{\sc iv}/Ly$\alpha$, He\,{\sc ii}~1640/Ly$\alpha$, Si\,\textsc{iv}/Mg\,\textsc{ii}&	  & \\				
NGC 4151&	H$\beta$/H$\alpha$, He\,{\sc ii}~4686/H$\alpha$, He\,{\sc i}~5876/H$\alpha$,&0.66	&17.9	&0	&51\%&19\%&28\%\\
 &C\,{\sc iv}/Ly$\alpha$, He\,{\sc ii}~1640/Ly$\alpha$, Si\,\textsc{iv}/Mg\,\textsc{ii}&	  &  \\		
 NGC 4593&	H$\beta$/H$\alpha$, He\,{\sc ii}~4686/H$\alpha$ &0.43	&19.1	&0	&40\%&22\%&29\%\\	
\enddata

\tablecomments{Column (1): Object name. Column (2): The fitting emission-line ratios. Column (3): The inner radius of BLR used in the model. Columns (4) and (5): The best-fitting minimum incident ionizing photon flux and coverage factor power-law index for radius. Columns (6), (7), and (8): The DC fraction at 3625~\AA, 5100~\AA, and 8100~\AA.}
\end{deluxetable*}

\section{Conclusions} \label{sec:conclusion}

We have constructed DC templates using \texttt{Cloudy} within the LOC model framework and performed spectral decomposition for a sample of 31 nearby type 1 AGNs ($z<0.24$) to constrain the contribution of the DC to the total continuum emission. The spectra are obtained from the STIS on the \textit{HST} and X-shooter on the VLT, providing wavelength coverage from the ultraviolet to the optical. Our main conclusions are summarized as follows:

1. We quantified the DC contribution using the flux ratios at ~3625~\AA, 5100~\AA, and 8100~\AA. For the full sample, whose rest-frame spectra cover at least the Balmer jump, the average DC fractions are \(0.33 \pm 0.14\) at 3625~\AA~ and \(0.13 \pm 0.08\) at 5100~\AA. These results demonstrate that the DC makes a substantial contribution to the AGN continuum emission (Section \ref{sec:decomp_results}; Figure \ref{DC_fraction}).

2. When fitting spectra that cover only a single discontinuity, the Balmer jump provides more reliable constraints on the DC component than the Paschen jump. 

3. Our mock spectral simulations demonstrate that the DC contribution can be reliably recovered with an appropriate template, whereas adopting an unsuitable template leads to systematic, wavelength-dependent biases in the inferred DC fraction. The decomposition results (including the DC templates) are publicly available at \url{https://doi.org/10.12149/101886}.

In the future, with the availability of higher-quality, higher-resolution, broader wavelength coverage, and time-domain observational data, we anticipate conducting more detailed, source-specific modeling, thereby placing tighter constraints on the physical conditions of the accretion disk and the BLR.

\begin{acknowledgments}
M.Q.F., S.Y.Z., and M.Y.S. acknowledge support from the National Natural Science Foundation of China (NSFC-125B2058, NSFC-12533006, NSFC-12322303), the National Key R\&D Program of China (No.~2023YFA1607903), and the Fundamental Research Funds for the Central Universities (20720240152). This work is also supported by the National Key R\&D Program of China (No.~2023YFA1607903, 2022YFF0503402). H. X. G. acknowledges support from the National Natural Science Foundation of China (NSFC, No.~12473018, 12522304) and the Overseas Center Platform Projects, CAS, No. 178GJHZ2023184MI. We acknowledge the AGN Variability Archive for providing the public catalogue. This research is based on observations made with the NASA/ESA Hubble Space Telescope obtained from the Space Telescope Science Institute, which is operated by the Association of Universities for Research in Astronomy, Inc., under NASA contract NAS 5–26555. The HST data used in this paper can be found in MAST at \url{https://doi.org/10.17909/57es-ve88}. 
\end{acknowledgments}

\software{astropy \citep{Astropy2013, Astropy2018, Astropy2022}, Matplotlib \citep{matplotlib}, Numpy \citep{2020NumPy-Array}, Scipy \citep{2020SciPy-NMeth}, Cloudy \citep{cloudy_2025}}, PyQSOFit \citep{pyqsofit_2018, Shen_2019}, TelFit,\footnote{https://telfit.readthedocs.io/en/latest/} spectool\footnote{https://github.com/zhang-zhixiang/spectool}. 

\facility{HST(STIS), VLT:Kueyen}

\newpage
\bibliography{reference}{}
\bibliographystyle{aasjournal}

\appendix
\restartappendixnumbering

\section{Disk Model Test}\label{disktest}
To evaluate the impact of different accretion disk models on the spectral decomposition, we performed a comparative analysis using Mrk~1048 as a representative example, for which the two power-law slopes in the BPL fit differ substantially. In addition to the broken power-law (BPL) model adopted in our main analysis, we apply a single power-law (PL) model and a standard disk (SSD) model, incorporating host-galaxy extinction effects for the latter.

For the SSD model, we compute the radial profile of the effective disk temperature using the black hole mass and the Eddington ratio. Following the standard thin-disk prescription \citep{Shakura_1973}, the effective temperature at radius $R$ is given by
\begin{equation}
T_{\rm eff}(R) = \left[ \frac{3GM_{\rm BH}\dot{M}}{8\pi \sigma R^{3}} \left( 1-\sqrt{\frac{R_{\rm in}}{R}} \right) \right]^{1/4},
\end{equation}
where $G$ is the gravitational constant, $\sigma$ is the Stefan-Boltzmann constant, and $R_{\mathrm{in}}$ is the inner radius of the accretion disk. The mass accretion rate $\dot{M}$ is derived from the Eddington ratio by assuming a standard radiative efficiency of $\eta = 0.1$. The parameters for each source, namely black hole mass ($M_{\mathrm{BH}}$) and the Eddington ratio, are listed in Table~\ref{tab:obs_t}. The disk is divided into concentric annuli, with each annulus treated as a local blackbody emitter at its effective temperature. The total monochromatic luminosity $L_{\nu}$ is then obtained by integrating over the disk surface:
\begin{equation}
L_{\nu} = \int_{R_{\rm in}}^{R_{\rm out}} 4\pi^{2}R\,B_{\nu}[T_{\rm eff}(R)]\,\mathrm{d}R,
\end{equation}
where $B_{\nu}$ is the Planck function. The resulting intrinsic spectrum is subsequently converted into the observed flux using the luminosity distance of the source.

To account for intrinsic dust extinction, we adopt the average Small Magellanic Cloud (SMC) extinction law, \texttt{G24\_SMCAvg} from \citet{Gordon_2024}, to the intrinsic disk spectrum. The fit includes two free parameters for the disk component: a global normalization and the color excess $E(B-V)$, with a physical constraint of $0 \leq E(B-V) < 1$.

\begin{figure*}
\includegraphics[width=1. \linewidth]{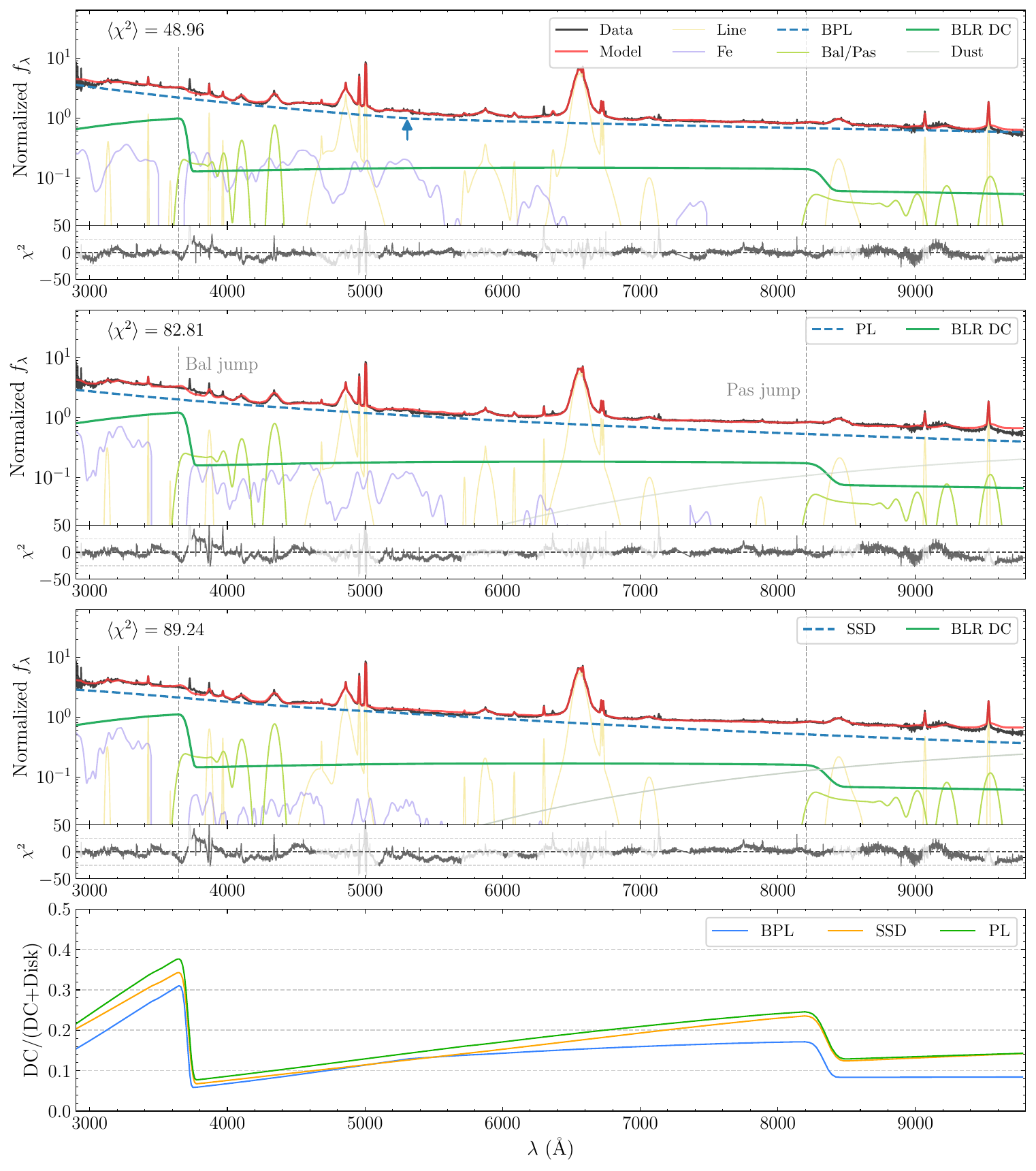}
\caption {Comparison of the DC contributions derived using three different accretion disk models for Mrk~1048. The upper three panels present the spectral decomposition results obtained with the BPL, PL, and SSD accretion disk models, respectively, with the corresponding $\chi^2$ values shown in the lower part of each panel. In the $\chi^2$ plots, dark curves mark the wavelength regions included in the continuum fitting, whereas light curves indicate the regions excluded from the fitting. In the bottom panel, the blue, green, and orange curves represent the DC fractions derived from the BPL, PL, and SSD decomposition results, respectively. These results illustrate the dependence of the derived DC fractions on the adopted disk model.}
\label{disk_text}
\end{figure*}

Figure~\ref{disk_text} presents the spectral decomposition results for Mrk~1048. Among the three models, the BPL model yields the lowest average reduced chi-squared value ($\sim 49$), indicating the best overall agreement with the observed spectrum, whereas the SSD model yields the highest value ($\sim 89$). The PL model yields the highest DC fractions. In contrast, the BPL model yields the lowest DC fraction around the Balmer jump ($\sim 30\%$), compared to $\sim 37\%$ for the PL model and $\sim 34\%$ for the SSD model. Similarly, around the Paschen jump, the DC fraction derived from the BPL model is $\sim 17\%$, whereas the PL and SSD models yield higher values of $\sim 24\%$ and $\sim 23\%$, respectively. Both the PL and SSD fits also require a substantial dust component. However, the SSD model fails to reproduce the observed luminosity-independent UV SED and the extremely red intrinsic EUV continuum of quasars \citep[e.g.,][]{Cai_2023}. The theoretical limitations might partly affect the DC contribution fitted with the SSD model.

\section{Other fitting results}\label{parameters}

Here we list other important fitting results in Table~\ref{tab:para}.

\begin{deluxetable*}{lccccccccc}
\tablecaption{Other important fitting results\label{tab:para}}
\tablehead{
\colhead{Object}&\colhead{$\lambda_{\mathrm{rest}}(\mathrm{\AA})$} &\colhead{DC FWHM (km s$^{-1}$)} &\colhead{$\alpha_\mathrm{short}$} &\colhead{$\alpha_\mathrm{long}$}& \colhead{$\lambda_\mathrm{break}~(\text{\AA})$ } & \colhead{$T_\mathrm{dust}~(\mathrm{K})$ } & \colhead{$R_\mathrm{dust} (8000\,{\text{\AA}})$ } & \colhead{$R_\mathrm{host} (5100\,\text{\AA})$} & \colhead{$R_\mathrm{host} (8000\,\text{\AA})$ }}
\colnumbers{}
\startdata
ESO~362-G18 & [2953, 10074]& 11460 & -1.66  &  0.00  & 3000 & 1509  &  0.01  & 0.54  &  0.36 \\
ESO~511-30 & [2923, 9971]& 3367 & -2.08  &  0.00  & 3349 & 2000  &  0.00  & 0.00  &  0.00 \\
Fairall~9 & [2863, 9753]& 4017 & -2.74  &  -0.84  & 4741 & 1072  &  0.00  & 0.00  &  0.00 \\
H~0507+164 & [2938, 10026]& 2778 & -0.08  &  -0.07  & 3000 & 1955  &  0.01  & 0.00  &  0.00 \\
Mrk~590 & [2912, 9937]& 6485 & -1.63  &  0.00  & 4238 & 1009  &  0.00  & 0.67  &  0.67 \\
Mrk~1044 & [2938, 10027]& 2057 & -2.15  &  -1.95  & 3478 & 1971  &  0.02  & 0.10  &  0.31 \\
Mrk~1048 & [2866, 9779]& 3976 & -2.14  &  -0.86  & 5307 & 1000  &  0.00  & 0.00  &  0.00 \\
Mrk~1239 & [2931, 10001]& 792 & -0.14  &  0.00  & 3000 & 1000  &  0.00  & 0.00  &  0.00 \\
Mrk~1298 & [2821, 9624]& 2645 & -1.62  &  -1.29  & 3268 & 1913  &  0.04  & 0.14  &  0.07 \\
NGC~2617 & [2948, 10057]& 2645 & -0.55  &  0.00  & 3477 & 1000  &  0.00  & 0.00  &  0.00 \\
NGC~3783 & [2963, 10109]& 4065 & -4.00  &  -1.28  & 4320 & 1000  &  0.00  & 0.00  &  0.00 \\
NGC~4748 & [2948, 10058]& 849 & -1.42  &  -0.32  & 3392 & 1000  &  0.00  & 0.00  &  0.00 \\
PG~2214+139 & [2806, 9573]& 4546 & -2.23  &  -0.96  & 4151 & 1997  &  0.00  & 0.00  &  0.00 \\
PG~0934+013 & [2848, 9712]& 1709 & -2.28  &  -1.20  & 4875 & 1001  &  0.00  & 0.00  &  0.00 \\
PKS~2004-447 & [2411, 8226]& 1421 & 0.00  &  0.00  & - & 1986  &  0.22  & 0.00  &  0.00 \\
3C~382 & [2740, 9687]& 18000 & -2.72  &  -1.75  & 4875 & 1000  &  0.00  & 0.00  &  0.00 \\
Arp~151 & [1115, 5581]& 2383 & -0.74  &  -0.74  & - & 1000  &  -  & 0.00  &  - \\
Mrk~50 & [1112, 5566]& 4996 & -1.71  &  -1.71  & - & 1000  &  -  & 0.00  &  - \\
Mrk~279 & [1105, 9932]& 9296 & -1.62  &  -1.62  & - & 1000  &  0.02  & 0.00  &  0.00 \\
Mrk~817 & [1104, 9940]& 5583 & -2.01  &  -2.01  & - & 1066  &  0.02  & 0.00  &  0.00 \\
Mrk~1310 & [1116, 5588]& 2385 & -1.48  &  -1.02  & 4907 & 1000  &  -  & 0.00  &  - \\
NGC~6814 & [1132, 5665]& 7500 & -0.78  &  -0.78  & - & 1000  &  -  & 0.00  &  - \\
PG~1226+023 & [2505, 8862]& 4797 & -2.25  &  -2.04  & 4315 & 1021  &  0.05  & 0.00  &  0.00 \\
PHL~1811 & [953, 4776]& 1426 & -1.60  & -1.60   & - & 1000  &  -  & 0.00  &  - \\
SBS~1116+583A & [1108, 5543]& 3507 & -1.43  &  -1.43  & -& 1000  & -  & 0.00  &  - \\
ZW~229-015 & [1107, 5543]& 2657 & -1.65  &  -1.65  & - & 1000  &  -  & 0.00  &  - \\
Mrk~110 & [1526, 9885]& 3233 & -2.03  &  -2.03  & -& 1010  &  0.02  & 0.00  &  0.00 \\
Mrk~493 & [1104, 9925]& 1495 & -1.49  &  -1.36  & 5187 & 1344  &  0.03  & 0.00  &  0.00 \\
Mrk~509 & [1100, 9887]& 4514 & -1.72  &  -1.72  & - & 1000  &  0.00  & 0.00  &  0.00 \\
NGC~4151 & [1135, 10236]& 7119 & -1.32  &  -1.09  & 4838 & 1912  &  0.02  & 0.00  &  0.00 \\
NGC~4593 & [2864, 10153]& 3499 & -3.32  &  -0.80  & 4918 & 1979  &  0.07  & 0.00  &  0.00 \\
\enddata

\tablecomments{ Column (1): Object name. Column (2): Wavelength coverage. Column (3): The full width at half maximum (FWHM) of the DC component, given to the nearest integer. Columns (4) and (5): The power-law indices $\alpha_\mathrm{short}$ and $\alpha_\mathrm{long}$ of the BPL model. Column (6): The break wavelength of the BPL model, given to the nearest integer. A dash (–) indicates that the break wavelength is not applicable because the two power-law slopes are identical, or that the break wavelength lies outside the spectral coverage. Column (7): Dust blackbody temperature. Columns (8)--(10): The dust fraction at 8000~\AA, and the host-galaxy contributions at 5100~\AA and 8000~\AA.}

\end{deluxetable*}

\end{document}